\documentclass[letterpaper]{aastex63}
\usepackage{natbib}
\usepackage{color}
\usepackage{array}  
\usepackage{verbatim}
\usepackage{amsmath}
\usepackage{brief_bib}
\usepackage{wasysym}
\usepackage{hyperref}

\usepackage[greek, english]{babel}
\usepackage{teubner}
\usepackage[T1,OT1]{fontenc}

\usepackage{lineno}

\usepackage{epstopdf}
\DeclareGraphicsExtensions{.ps}
\newcommand{\zz}{``{\sl\koppa}''} 
\newcommand{\zzs}{``{\sl\koppa}'' } 
\renewcommand{\refname}{} 
\renewcommand{\bibname}{}
\newcommand\p{\\}
\def\eqref#1{Equation~(\ref{#1})}
\def\figref#1{Figure~(\ref{#1})}
\def\secref#1{Section~\ref{#1}}
\newcommand\vi{\epsilon}
\newcommand\rvec{{\bf r}}

\newcommand\yhat{{\bf \hat{y}}}

\newcommand\shat{{\bf \hat{s}}}
\newcommand\uhat{{\bf \hat{u}}}

\def\mm{MURaM}

\definecolor{gray}{RGB}{180, 180, 180}
\definecolor{dblue}{RGB}{0, 0, 100}

\newcommand{\nv}[2]{#2} 

\begin{document}
\title{The Coronal Veil}
\author{A.~Malanushenko$^{1}$, M.C.M.~Cheung$^{2}$, C.E.~DeForest$^{3}$, J.A.~Klimchuk$^{4}$, M.~Rempel$^{1}$}
\affil{$^1$High Altitude Observatory, National Center for Atmospheric Research, Boulder, CO, USA\\
       $^2$Lockheed Martin Solar and Astrophysics Laboratory, Palo Alto, CA, USA\\
			 $^3$Southwest Research Institute, Boulder, CO, USA\\
			 $^4$Heliophysics Division, NASA Goddard Space Flight Center, Greenbelt, MD, USA}

\begin{abstract}
Coronal loops, seen in solar coronal images, are believed to represent emission from magnetic flux tubes with compact cross-sections. We examine the 3D structure of plasma above an active region in a radiative magnetohydrodynamic simulation to locate volume counterparts for coronal loops. In many cases, a loop cannot be linked to an individual thin strand in the volume. While many thin loops are present in the synthetic images, the bright structures in the volume are fewer, and of complex shape. We demonstrate that this complexity can form impressions of thin bright loops, even in the absence of thin bright plasma strands. We demonstrate the difficulty of discerning from observations whether a particular loop corresponds to a strand in the volume, or a projection artifact. We demonstrate how apparently isolated loops could deceive observers, even when observations from multiple viewing angles are available.\p

While we base our analysis on a simulation, the main findings are independent from a particular simulation setup and illustrate the intrinsic complexity involved in interpreting observations resulting from line-of-sight integration in an optically thin plasma.\p

We propose alternative interpretation for strands seen in EUV images of the corona. The ``coronal veil'' hypothesis is mathematically more generic, and naturally explains properties of loops that are difficult to address otherwise---such as their constant cross section and anomalously high density scale height. We challenge the paradigm of coronal loops as thin magnetic flux tubes, offering new understanding of solar corona and, by extension, of other magnetically confined bright, hot plasmas.
\end{abstract}

\section{Introduction}\label{sec_intro}
Coronal loops are thin emitting strands seen in the solar atmosphere (called corona), which are most often observed in X-Rays and in Extreme Ultraviolet (EUV), suggesting the plasma in them is heated to millions of degrees. They generally appear as curved strands, tens to hundreds megametres long, with major diameters hundreds to thousands of kilometers; bundles of loops are often seen connecting pairs of sunspots with opposite polarity of magnetic field at the solar surface \cite{Reale2010}. \p

Loops are usually associated with thin magnetic flux tubes, as they seem to take shapes anticipated for magnetic field lines, both in the large-scale corona \cite{Schrijver2010} and in cores of active regions \cite{Wiegelmann2012}. This resemblance in shape usually allows researchers to interpret coronal loops as plane-of-sky projections of thin magnetic flux tubes which are filled with plasma with different properties (e.g., hotter/cooler/denser) than plasma in the surrounding corona. The region between these thin flux tubes is also filled with magnetic field, but the plasma in it is fainter because it has a lower density and/or different temperature to which the observations are less sensitive. The physical mechanisms that are responsible for heating and mass input into the corona is still a topic of active research \citep{Klimchuk2015}. However, once heat is produced, it is conducted along the magnetic field much more efficiently than across it \citep{Ireland1992}, and once plasma is injected into the corona, it is normally bound to move along the magnetic field lines \citep[so-called ``frozen-in'' condition, common in astrophysical plasmas---e.g.,][]{Priest2000, Longcope2005b}. \p

While coronal loops themselves have been known since the 1960's, some of their observed properties remain challenging to explain. Firstly, active region (AR) coronal loops exhibit scale heights much taller than expected from gravitational scale height calculations \citep[e.g.,][]{Doyle1985, Aschwanden2000, Winebarger2003, Fuentes2006}. Observed flows \citep[e.g.,][]{Fredvik2002, UgarteUrra2009} are not sufficient to support the plasma at such heights via ballistic or siphon momentum transfer \citep{DeForest2007}. Waves have been observed with many different instruments \citep[e.g.,][and references therein]{Winebarger2003}, but they do not provide sufficient wave pressure to support the material. This leaves the question of why the observed scale height is so much higher than the inferred one.\p

Secondly, coronal loops appear to lack expected visual expansion with height, as the confining magnetic field weakens with altitude on average in the corona. This observation extends through every extreme ultraviolet (EUV) imager to date, from Skylab's SO82A ``Overlappograph'' \citep{Tousey1977, Feldman1987} to current instruments such as SDO/AIA \citep{AIA_ref} and the Hi-C rocket \citep{Kobayashi2014}. The fixed minimum apparent width varies by instrument \citep{DeForest2007} but has been shown to be ``real'' in the sense it is present in the data \citep{Fuentes2006, Fuentes2008}. The behavior appears consistent with unresolved isotropic or resolved anisotropic tapering in confined loop structures \citep{DeForest2007, Malanushenko2013, Peter2013, Chastain2017} and has been used to explain loop scale heights \citep{Martens2010, Scott2012}. However, a separate line of work \citep[e.g.,][]{Brooks2012, Dudik2014, Aschwanden2013} has developed, under the hypothesis that the physical structures in the corona are resolved and therefore reflect the actual coronal geometry.\p

\citet{Malanushenko2013} had pointed out that anisotropic expansion of magnetic flux tubes may lead to an observers bias: should loops be tapered structures, those that expand along the line of sight (and, consequently, appearing brighter and thinner) would predominantly be selected for analysis. This work was followed by several studies aimed to confirm or reject this hypothesis observationally using various techniques to gather loops' aspect ratio \citep{Kucera2019, Klimchuk2020, McCarthy2021}. All three works reported mixed results. \citet{Kucera2019} is the first of this kind and highlights the techniques and challenges of this analysis. They found that for 2 loops, the hypothesis of anisotropic expansion could not have been confirmed or ruled out; they concluded that the results were largely consistent with nonexpanding or weakly expanding loops. Further, \citet{Klimchuk2020} looked at loops observed in 193\AA{} by Hi-C and examined, along each loop, its peak intensity \textit{vs.} standard-deviation width, after background subtraction. For a tapered structure of constant intensity which is twisting about its axis, there would be a negative correlation between these two quantities. \citet{Klimchuk2020} found that out of 22 loops, 4 were consistent with a negative correlation, 11 loops were consistent with a positive correlation, and 7 exhibited no significant correlation between width and intensity. \citet{McCarthy2021} looked at an AR system observed in 171\AA{} in quadrature by AIA and STEREO. They found that for 151 loops observed by AIA which could be fit to magnetic field lines, about 70\% had possible counterparts in STEREO observations. For each loop for which in-depth analysis was feasible, two correlations were studied: (a) between diameters as observed from two viewing angles, as well as (b) between diameter and intensity observed in the same viewing angle. They found that out of 13 loops suitable for the analysis, 4 were consistent with elliptical cross-sections in the sense that both (a) and (b) correlations were negative, 4 exhibited positive (a) and (b) correlations which is consistent with circular cross-section more than with elliptical one, and 5 exhibited mixed behavior. It is worth noting that \citet{Klimchuk2020} was focused on starting segments or short loops in a large AR with complex structure, while \citet{McCarthy2021} studied near-full-arches in dipolar ARs, so the lengths of studied loops and the AR configuration were significantly different \citep{McCarthy2021}.\p


In the present work, we explore a hypothesis different from all of the above: that the individual structures are in fact \textit{bundles} of coronal loops; they might be extended and complex structures, manifesting as diffuse emission with an overlay of apparently constant-width, large scale-height loops, and that individual loops only appear to be compact monolitic structures, but many are in fact projection artifacts. While it is not currently possible to sample the corona in 3D directly, \textit{ab initio} numerical simulations have developed sufficiently to explore realistic coronal loop systems and determine whether this geometric hypothesis is plausible.\p

\nv{}{We note that the concept of thin veil-like sheets that are responsible for apparent strand-like compact features is \textit{per se} not new in astrophysics. For example, \citet{Hester1987} described filamentary remnants of a supernova such as the Cygnus Loop and IC 443 and argued that the observed structure is best explained as optical artifacts from a corrugated veil-like surface, as opposed to a set of rope-like or cloud-like structures. For another example, some solar polar plumes are consistent with projections of curtain-like structures \citep[e.g.,][]{Wang1995, Gabriel2003, Gabriel2009}. Further, \citet{Koutchmy2001} had studied a reconstruction of a global coronal field and concluded that the helmet structures observed in the global corona are consistent with folds in a surface of the reversal of radial field. In solar physics of smaller scales, \citet{Judge2011} proposed a sheet-like interpretation for solar spicules that can explain their apparent supersonic velocities. Finally, \citet{Howard2017} have argued that a three-part structure of CMEs can be explained as a optical artifact, produced by caustics at the edges of a thin hollow sheet around a kinking flux rope.}\p

The further paper is organized as follows. In \secref{sec_methods}, we describe and present 3D simulations of AR loop system performed with a new version of the radiative magnetohydrodynamic simulator ``MURaM'', which produces realistic magnetoconvection and has recently been extended to allow simulation of the corona \citep{Rempel2017}. The simulations include full forward modeling of optically thin coronal EUV emission including detection by a simulated instrument; we qualitatively show that the simulations reproduce the generic behavior of active region loops on the Sun. This behavior includes the tall scale apparent height and apparent constant-width features as well as other puzzling aspects, such as difficulty of end-to-end loop tracing, that have been hard to reproduce with models of loops as confined structures.\p

In Sections \ref{sec_analysis}-\ref{sec_single_loop}, we explore the simulation in more detail, including case studies of particular features as seen through the simulated instrument and in cross sections of the simulation data volume. We show (A) that although the simulated instrument observes fine-scale apparently-confined structures in the image plane, there are often no corresponding fine-scale fully confined (compact) structures in the cross section; (B) that in the datacubes shown in the paper there is often little correspondence between the brightest structures in the simulated corona and the bright fine structures in the simulated observation; (C) that the bright structures appear to be aligned along magnetic field lines, and (D) while quantitative statistics of loops' volumetric counterparts is outside of scope of this study, we nonetheless demonstrate in Appendix~\ref{sec_appendix} that many of the bright fine structures in the simulated observation correspond to areas of near tangency between the line of sight and extended veil-like manifolds. In \secref{sec_stereo} we discuss difficulties arising when trying to apply stereoscopic techniques to veil-like manifolds.\p

In \secref{sec_results} we summarize our findings, and in \secref{sec_other_mhd} discuss their generality and their robustness in light of current understanding of active region loops. We suggest that the complex cross-section described in \secref{sec_single_loop} (the ``veil'') is more mathematically general than that of thin monolitic loops considered in most work to date, and infer that it is therefore plausible as an explanation for the currently observed characteristics for coronal loops. We note that since observationally it may be difficult to tell whether a particular observed structure is a veil or a bundle of strands, the former must be ruled out if measurements which assume the latter are to be made. \p

In \secref{sec_conclusions}, we summarize our conclusions and their importance for thecurrent understanding of coronal loops, including geometric structure, plasma characteristics, and potential heating mechanisms. \p

\section{Methodology}\label{sec_methods}

For this work we used the MuRaM MHD simulation code (\secref{sec_muram}) with a forward emission and instrument model (\secref{sec_making_emission}) to compare `` ground truth'' of a particular simulated solar corona to a simulated instrument output when that corona is observed remotely. The forward modeled results permit plausibility checks by direct comparison to data from real solar instrumentation.

\subsection{MHD Simulation}\label{sec_muram}

In this work, we use simulations of an active region created with the MURaM code. MURaM is a 3D radiative MHD code \citep{Voegler2005} which has been recently extended into the corona by \citet{Rempel2017}. The code solves the single fluid MHD equations in a domain ranging from the upper convection zone into the lower corona. The equation of state considers equilibrium ionization for a solar element mixture through a combination of OPAL \citep{Rogers1996} in the convection zone and the Uppsala Opacity Package \citep{Gustafsson1975} in chromosphere and corona. The energy equation accounts for radiative heating and cooling, which is computed in photosphere and chromosphere through a full 3D radiation treatment assuming local thermal equilibrium and in the corona through optically thin loss function based on CHIANTI \citep{CHIANTI_v7}. In the corona the code considers in addition magnetic field aligned Spitzer heat conduction. The numerical treatment of the latter is based on the hyperbolic approach detailed in \citet{Rempel2017} that prevents severe time-step constraints in a fully explicit treatment by filtering out unphysical transport velocities in regions with high coronal temperatures. Furthermore, the Boris correction \citep{Gombosi2002, Rempel2017} is used to artificially limit Alfv{\'e}n velocities and prevent severe time-step constraints. Within the context of these simulations the corona is self-maintained. Magneto-convection in the photosphere does lead to an upward directed Poynting flux that maintains a corona at a temperature of 1-2~MK through a combination of numerical viscous and resistive heating. \nv{}{The side boundaries are periodic, and the top boundary is open to mass flow but is set to dampen the vertical flows and to have zero Poynting flux \citep[see][for more details]{Rempel2017}.}\p

In particular, we examine the simulation featured in \citet{Cheung2019} of a flux emergence near a pre-existing magnetic dipole. \citet{Cheung2019} reported a C-class flare associated with magnetic reconnection between the two dipoles, and demonstrated a wealth of phenomena commonly observed in flares. \p

\begin{figure}[h] 
 \begin{center} 
  \includegraphics[width=17cm]{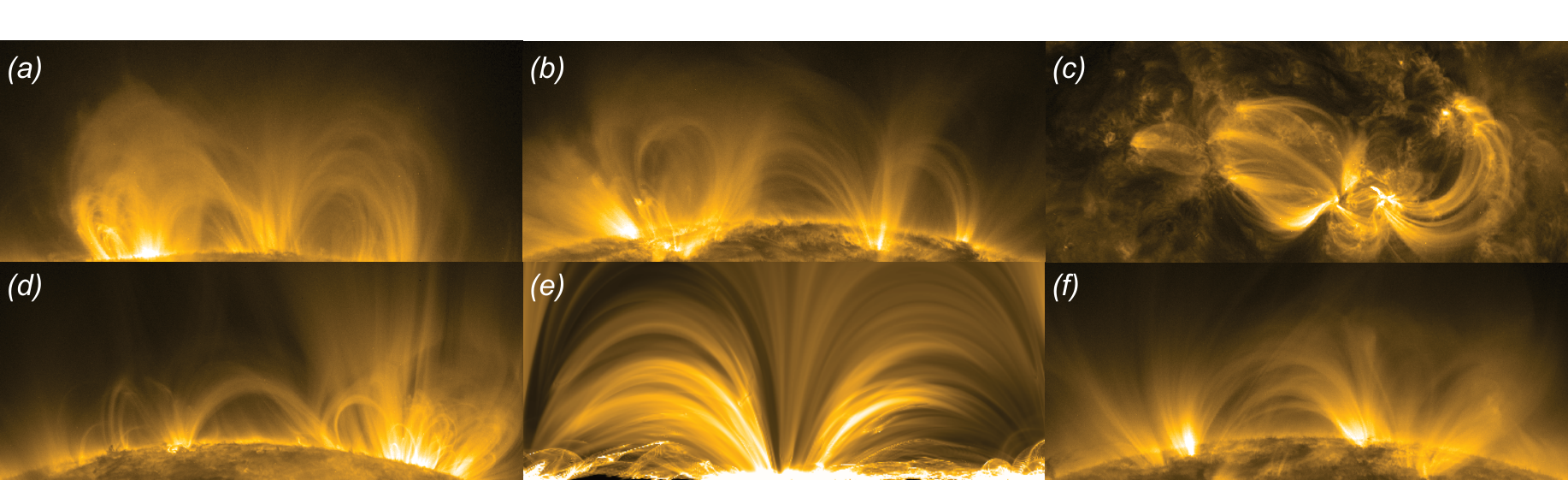}
 \end{center} 
 \caption{Extreme Ultraviolet (EUV) images of five coronal active regions in 171\AA~ as seen by AIA (panels a-d and f) are arrayed around the synthetic observation analyzed in this paper (panel e). The images are false-color and the color scale has increments proportional to the root of intensity.  Each image is drawn in a different intensity range, chosen to highlight the relevant features in that image.  The fields of view also vary, chosen for each image individually to present a particular relevant example structure. \nv{}{Note that we selected regions with loop bundles oriented in such a way that roughly matches the viewing angle of the simulation in panel e.)}}	
 \label{aia_panels}
\end{figure}

\begin{figure}[h] 
 \begin{center} 
  \includegraphics[width=17cm]{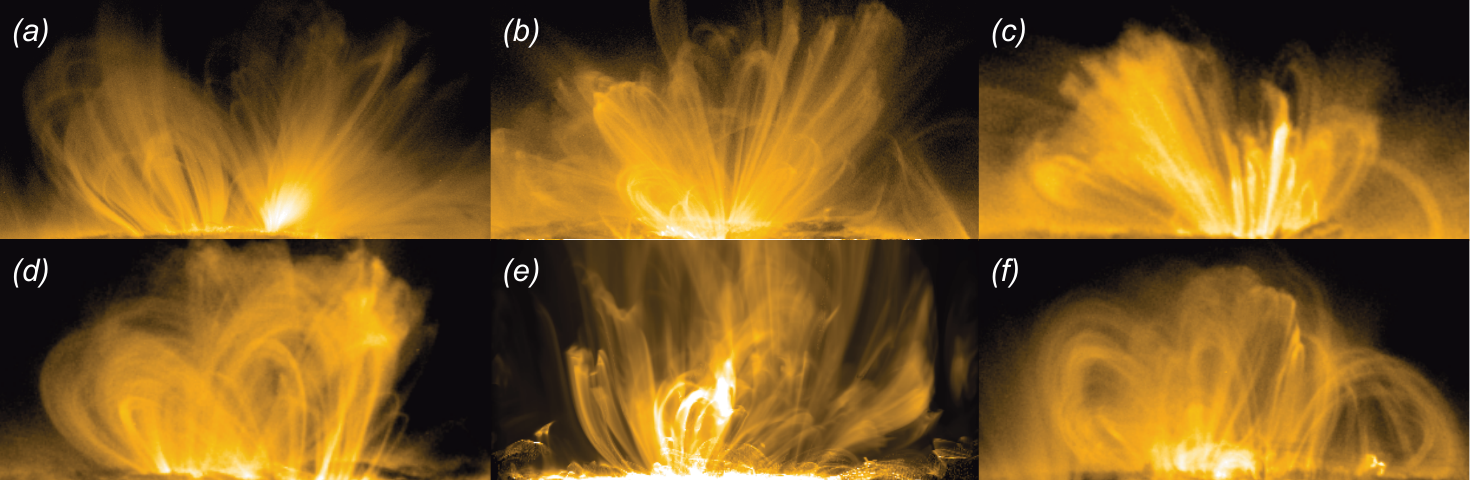}  
 \end{center} 
 \caption{\nv{}{The same as \figref{aia_panels}, but with a set of different ARs viewed from the limb (panels a-d and f) around the synthetic observation (panel e) rotated to roughly $70\degr$ with respect to the viewing angle in \figref{aia_panels}.}}	
 \label{aia_panels_limb}
\end{figure}

In \nv{\figref{aia_panels}}{Figures~\ref{aia_panels} and~\ref{aia_panels_limb}}, we compare synthesized observations from this simulation with real data (we explain how they were synthesized further in \secref{sec_making_emission}). As this is an \textit{ab initio} simulation, there is no single active region to compare it \textit{to} (although the general configuration of the dipoles has been \textit{inspired} by an active region NOAA 12017, observed in March-April 2014, only its general properties are used, such as presence of parasitic flux emergence near one of the existing \nv{ARs}{polarities}). We therefore choose to not compare the synthetic data to one observation, but rather show \nv{it}{them} in context of many active regions. Our aim is to give readers a feel of how EUV observations of active region coronae may appear, in a particular SDO/AIA wavelength channel. \p

\begin{figure}[h] 
 \begin{center} 
   \includegraphics[width=17cm]{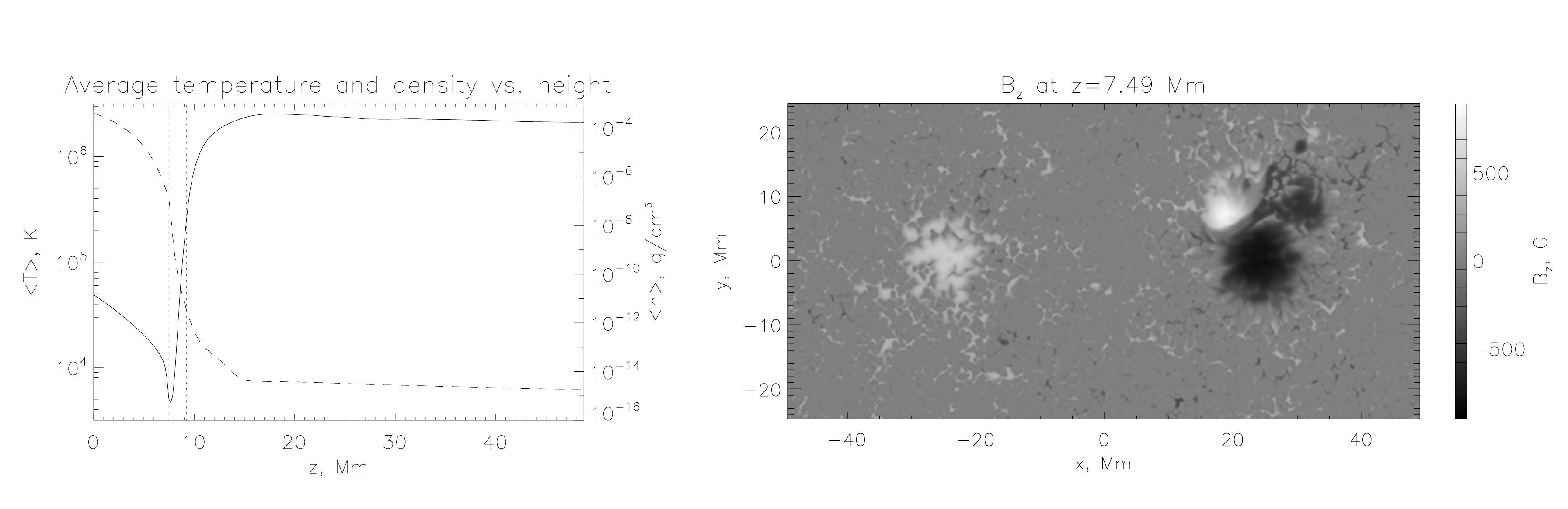} 
 \end{center} 
 \caption{\textit{(Left panel)} temperature (solid line) and density (dashed line) averaged over horizontal slices; two vertical lines indicate approximate location of the photosphere and the location of the base of the subvolume which we further study. \textit{(Right panel)} $B_z$ at a horizontal slice which approximately corresponds to photosphere.}	
 \label{muram_info1}
\end{figure}

We note that active regions are small compared to the Sun, and adopt local Cartesian coordinates, with $x$ and $y$ being local horizontal directions (E/W and N/S, respectively) and $z$ being vertical, as in \figref{muram_info1}. The line-of-sight (LOS) direction is usually $\yhat$. This coordinate system is used throughout the following sections. We stress this definition now as many of the further figures include complex 3D views, panels, slices, and line-of-sight (LOS) integrals, and the common coordinate system helps understand their spatial relationship. \p

\begin{figure}[h] 
 \begin{center} 
   \includegraphics[width=17cm]{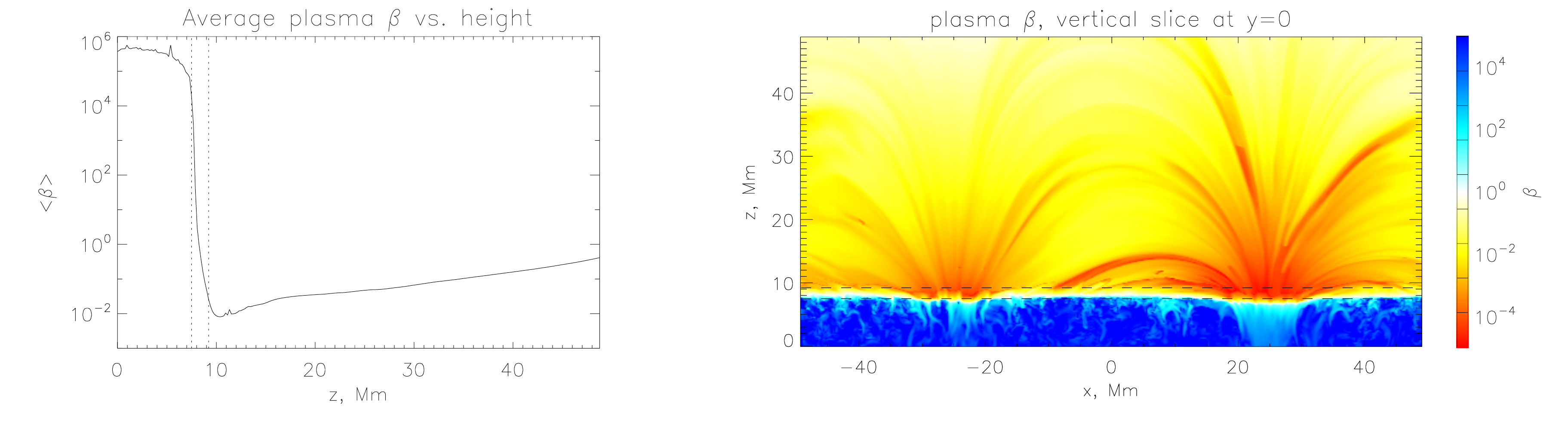} 
 \end{center} 
 \caption{\textit{(Left panel)} plasma $\beta$ averaged over horizontal slices; the vertical lines are as in \figref{muram_info1}, left panel. \textit{(Right panel)} $\beta$ at a vertical slice.}	
 \label{muram_info2}
\end{figure}

The simulation domain spans $98.304\times 49.152\times 49.152\mbox{ Mm}$ with a uniform grid spacing of $192\times 192\times 64\mbox{ km}$. In the vertical direction the domain spans from about $7.5\mbox{ Mm}$ beneath the photosphere to $42\mbox{ Mm}$ into the corona. We focus here on a snapshot from this simulation prior to the occurrence of the C-flare. The coronal magnetic field is a combination of a mostly potential active region with a newly emerged strongly twisted dipole near one of the sunspots. Figures~\ref{muram_info1} and~\ref{muram_info2} give a brief introduction to the volume: horizontal averages of density, temperature and plasma~$\beta$, a horizontal slice of $B_z$ at an approximately photospheric location ($z\approx 7.49\mbox{ Mm}$), and also a vertical slice of plasma~$\beta$ at the middle ($y=0$) of the domain. We further focus on the coronal portion of the volume, $z\geq 9.2\mbox{ Mm}$. \p

\subsection{Synthesizing Emission}\label{sec_making_emission}
We synthesize the emission using the CHIANTI database of atomic parameters \citep{CHIANTI_v7} and an SDO/AIA instrument response functions \citep{AIAref}. The intensity of the light, as registered by a particular instrument, emitted by locally isothermal plasma at a given location (we will further refer to it as \textit{volumetric emissivity}) in a given wavelength $\lambda$ is proportional to 
\begin{equation}
\vi(\rvec, \lambda)=n^2(\rvec)R(\lambda, T(\rvec)), 
\label{eqn_dI}
\end{equation}
where $n(\rvec)$ is plasma density, $T(\rvec)$ is its temperature, and $R(\lambda, T)$ is the temperature-response function, which includes both the atomic physics model and the response of a given instrument in a given wavelength. Further, in optically thin plasmas $\vi(\rvec)$ must be integrated along the line of sight (LOS) to obtain the observed intensity. Assuming $\yhat$ to be a LOS direction and $(x, z)$ to be plane-of-sky (POS), we calculate intensity at a given POS location as 
\begin{equation}
I(x, z)=\int\limits_{y=0}^{\infty}{\vi(\rvec)dy}. 
\label{eqn_I}
\end{equation}

In literature, the integral in \eqref{eqn_I} over LOS is typically rewritten, in the manner of LeBesgue integration, into an integral over temperature rather than physical distance using a quantity called differential emission measure (DEM):
\begin{equation}
I(x, z)=\int\limits_{T=0}^{\infty}{\mbox{DEM}(T)R(\lambda, T)dT}. 
\label{eqn_I_DEM}
\end{equation}
DEM reflects the emission of \textit{all} plasma along the line of sight at a given temperature, and it is useful for analyzing thermal composition of plasma \citep{Aschwanden_book}. \p

In the present work, however, we are interested in emitting structures \textit{in the volume}. We therefore calculate volumetric emissivity $\vi$ in the volume and use \eqref{eqn_I} to synthesize the observables. \p 

\section{Emission --- in the plane of sky and in the volume}\label{sec_analysis}

\begin{figure}[h] 
  \begin{center} 
   \hspace{-0.75cm}\includegraphics[width=18.75cm]{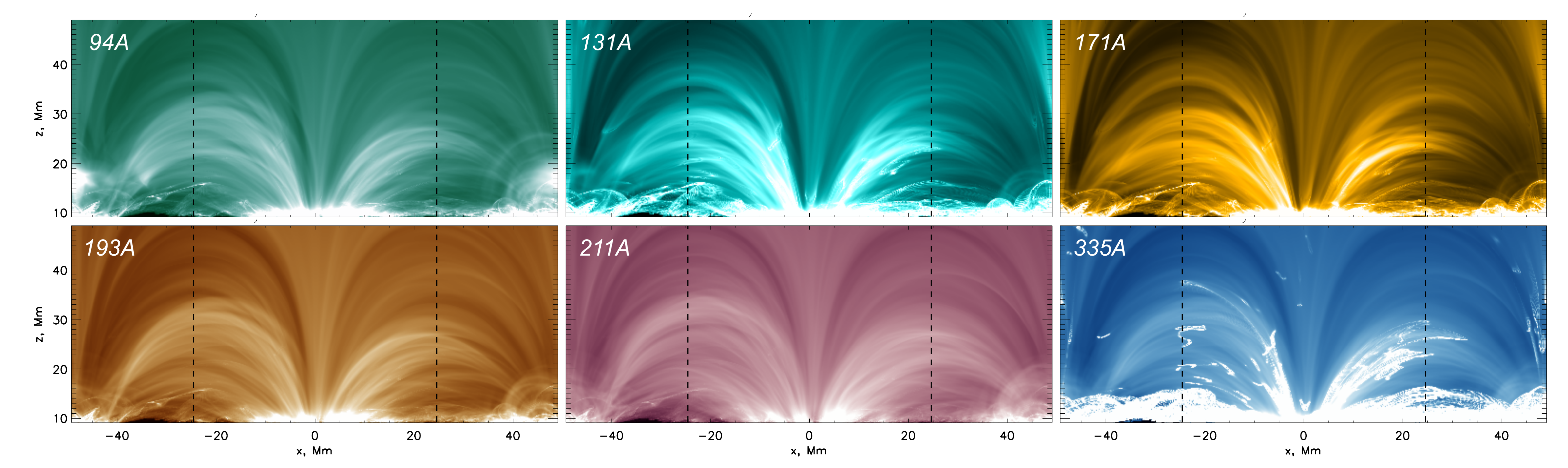}  
  \end{center} 
 \caption{Synthetic emission as it would appear in six extreme ultraviolet (EUV) channels of SDO/AIA. Line-of-sight (LOS) is along $y$ direction. We use the fact that the domain is periodic in $x$ and $y$ and shift the images in $x$ by $\approx 25\mbox{ Mm}$ (compared to Figures~\ref{muram_info1} and~\ref{muram_info2}) to better display both loop bundles. Vertical lines indicate two $(y,z)$ planes which we further examine in detail. All EUV images here and further are shown in a square-root intensity scale.} 
 \label{synth_aia_panels}
\end{figure}

\figref{synth_aia_panels} shows synthetic emission in six AIA channels, assuming $\yhat$ to be LOS direction. The simulation is periodic in horizontal directions and a pair of the two major sunspots from \figref{muram_info1} produces two bundles of loops (we hereafter shift the domain in $x$ by $\approx 25\mbox{ Mm}$ to have uninterrupted display of both bundles). \p

We find that in all channels, there are two large bundles of loops connecting the two sunspots. Each of these bundles has at least a dozen coronal loops, out of which many do not notably expand with height upon visual examination, which is qualitatively consistent with observations \citep[e.g.,][and references therein]{Reale2010}. The thinnest loops going down to the image resolution, which is also consistent with observations \citep{HiC_ref}. Additionally, there seem to be many loops visible at the apices of the bundles, which is not inconsistent with pressure scale height being larger than the height of the volume ($\approx 50\mbox{Mm}$), which is also consistent with current models \citep{Aschwanden_book}.\p 

The standard interpretation of these observations is that in the volume, there are two groups of thin (compared to their length) magnetic flux tubes filled with plasma of such a temperature as to warrant emission in AIA channels. These individual flux tubes overlap along the line of sight (LOS) making an impression of loop bundles, yet the common expectation is that the individual tubes are spatially isolated from one another in the volume. Consequently, the presence of several dozens of observed loops in the synthesized image implies the existence of several dozens of distinct flux tubes in the volume, and the diameters of these tubes are comparable to the observed widths of the loops. \p

We examine next two vertical slices on the synthetic images in \figref{synth_aia_panels}, passing roughly through the apices of loops the two loop bundles. The image intensity on each slice is a result of a LOS-integration of a plane in the volume: $I(x_0, z)=\sum{\vi(x_0, y, z)\Delta y}$. In line with the previous paragraphs, if a certain amount of thin strands is seen crossing this vertical line on the synthetic image, then the distribution of $\vi$ in the corresponding $(y, z)$ plane in the volume must contain comparable amount of small bright blobs. These blobs would be spatially separated, which is caused by spatially separated heating events (so the strands remain distinct); their diameters should be comparable to the thicknesses of the strands; and their number should be comparable to the number of strands observed on the synthetic image. \p

Figures~\ref{synth_midslices_x_128} and~\ref{synth_midslices_x_384} show volumetric emissivity in two vertical planes which project to the two dashed lines in \figref{synth_aia_panels}. It is evident that structures in these images can not be easily described as a set of numerous, small, compact blobs which in turn correspond to individual coronal loops. Instead, most of these images contain large-scale structures of complex shape, with numerous ridges and wrinkles. These structures are not, with few exceptions, easy to separate from one another. While some loops could be mapped to distinct, bright blobs, many loops do not seem to have a clear correspondence with the isolated structures in the volume. \p

We further argue that the brightest blobs alone are not sufficient to explain the numerous coronal loops in this simulation. We also demonstrate how the wrinkles in the large-scale structures may cause \textit{impressions} of loops. However, the immediate next step is to determine whether the structures seen in these two slices are extended in space, and whether they follow magnetic field lines. \p

\begin{figure}[h]
  \begin{center} 
   \includegraphics[width=10.8cm]{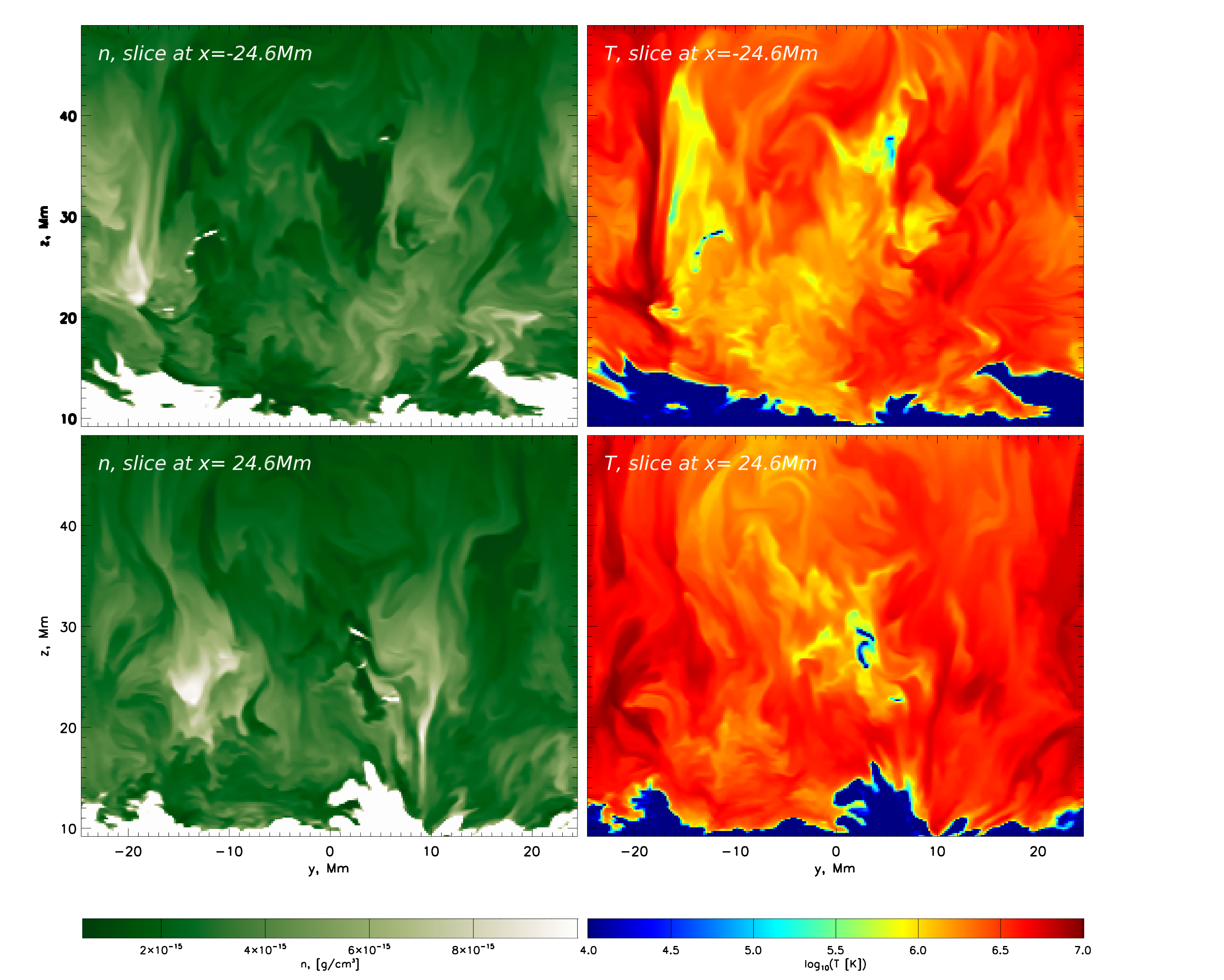}  
  \end{center} 
 \caption{Density (left panels) and temperature (right panels) in a $(y, z)$ plane for $x=\pm 26.4\mbox{ Mm}$ (the slices indicated in \figref{synth_aia_panels} by the vertical dashed lines).} 
 \label{midslices_t_rho}
\end{figure}

\begin{figure}[h] 
  \begin{center} 
   \includegraphics[width=18cm]{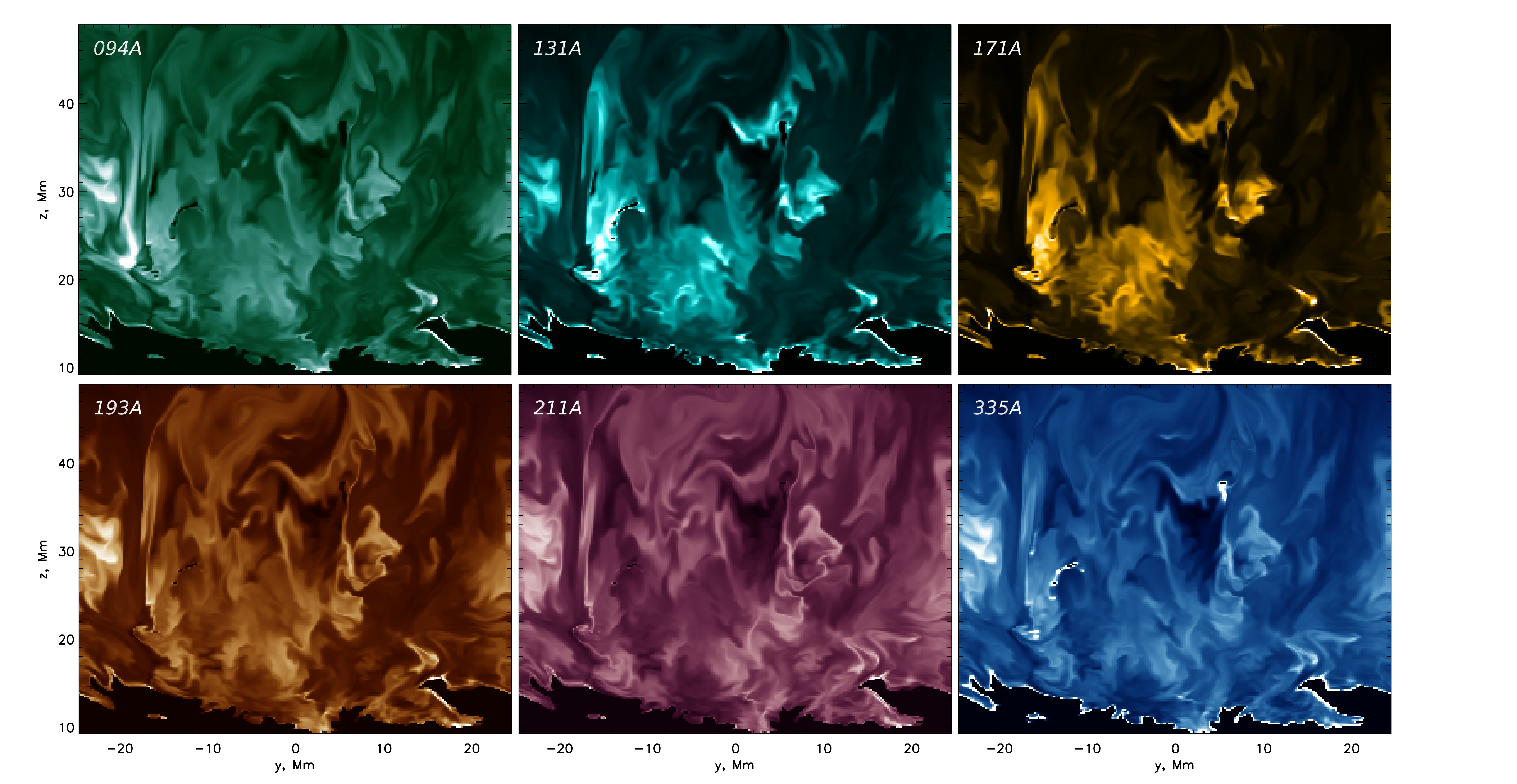}  
  \end{center} 
 \caption{Volumetric emission in a $(y, z)$ plane for $x=-24.6\mbox{ Mm}$ in six AIA channels. The $y$-integral in this plane projects into a single column of synthetic AIA images in \figref{synth_aia_panels}, indicated in that figure by one of the vertical dashed lines. All volumetric emission here and further is shown in a square-root intensity scale.} 
 \label{synth_midslices_x_128}
\end{figure}

\begin{figure}[h]
  \begin{center} 
   \includegraphics[width=18cm]{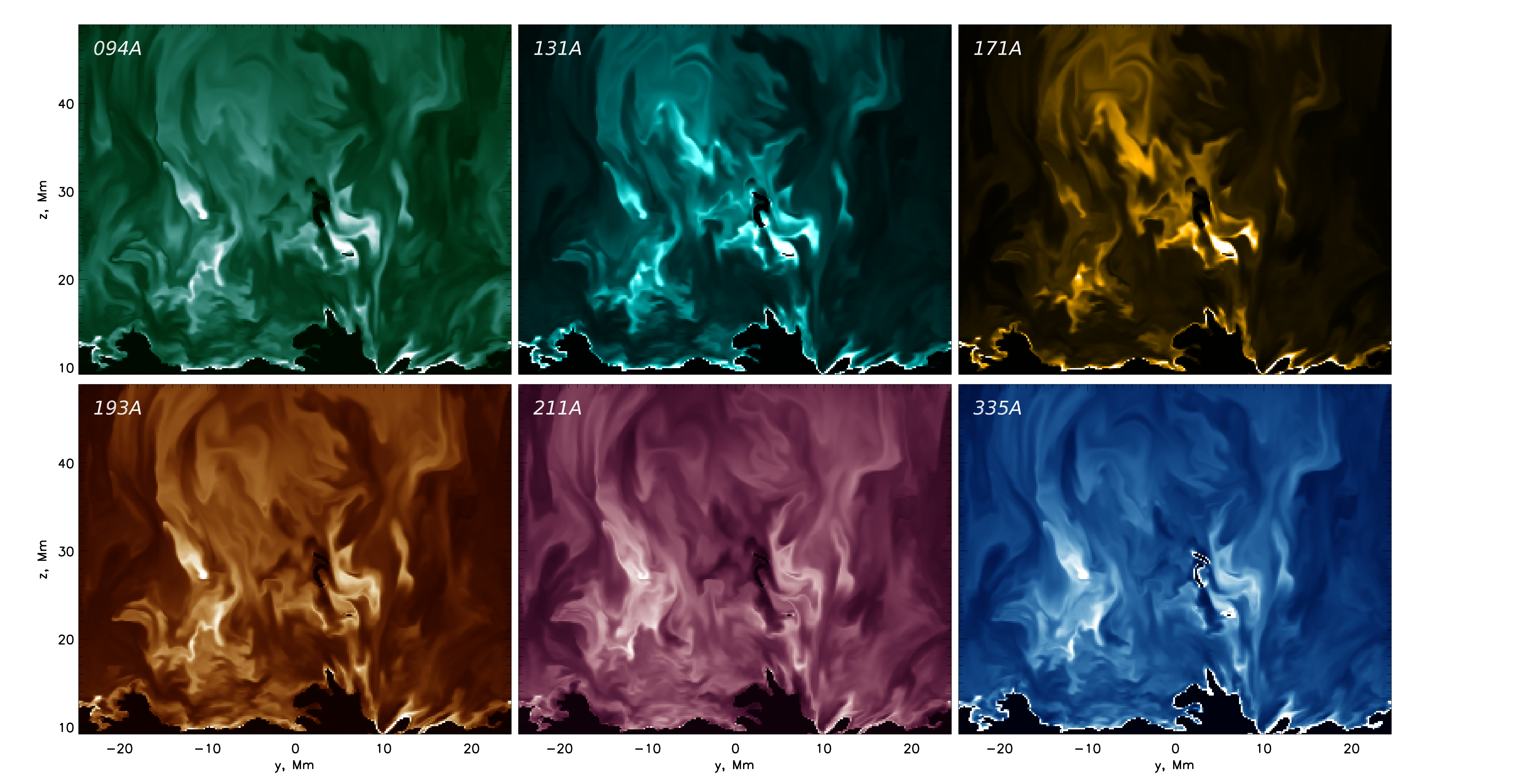}  
  \end{center} 
 \caption{Volumetric emission in a $(y, z)$ plane for $x=26.4\mbox{ Mm}$ in six AIA channels. The $y$-integral in this plane projects into a single column of synthetic AIA images in \figref{synth_aia_panels}, indicated in that figure by one of the vertical dashed lines.} 
 \label{synth_midslices_x_384}
\end{figure}

\section{Magnetic connectivity}\label{sec_magn_connectivity}

Much current work suggests or assumes that emitting plasma structures are confined within individual, compact magnetic flux tubes \citep[e.g.][]{Aschwanden2013, Klimchuk2015, Schmelz2006}. This assumption plausibly follows from the frozen-in magnetic field configuration \citep[e.g.,][]{Priest_book} in the local absence of topologic divides. This means that a localized density enhancement may propagate along, but not across the magnetic flux tube which embeds such enhancement; also, a localized temperature enhancement is well conducted along the embedding magnetic flux tube, but not across it. \p

We note that there are theories which suggest that ``loops'', in the sense of them being \textit{``bright elongated structures in the plane of sky''}, may not identically correspond to specific magnetic flux tubes. The examples include \citet{Peter2012}, who point out that temperature variation along an individual flux tube may be sufficient to place some regions of the flux tube outside of the thermal band which contributes most to a channel in which an observation is made. In this case, a combination of nearby flux tubes with slightly different temperatures/densities may cause an impression of a loop which in fact crosses several flux tubes. \citet{Fengs_thesis} have examined time evolution of a magnetic flux tube filled with emitting plasma in a 3D MHD simulation of an AR. They found that even when there is a correspondence between a compact flux tube (embedding a compact emissivity structure) and the appearance of a loop at one point in time, the subsequent evolution breaks compact emissivity structure into fragments, significantly distorts the magnetic flux tube's cross-section, and reduces the association between the evolved flux tube and the evolved loop. They also confirmed previous \citet{Peter2012} inference that high density and temperature structures are not cospatial, and different gradients of both within the magnetic flux tube they studied lead to nonuniform contribution to the appearance of a loop from different portions of the same flux tube at different times. Another example is \citet{DeForest2007}, who points out that individual emitting flux tubes could be much smaller than the resolution of an instrument, and they could spread apart with height as expected from magnetic field lines. In this case, the line-of-sight integral of a bundle of such unresolved features may cause an impression of a loop along the middle of such a bundle, where most features overlap. \p

As we showed in the previous section, the brightness enhancements in the volume of our simulation are not in general compact. But are these shapes confined to magnetic flux tubes, abeit of complicated cross-section? \p

\begin{figure}[h]
  \begin{center} 
\begin{interactive}{animation}{movie_211a.mp4}
  \includegraphics[height=9.5cm]{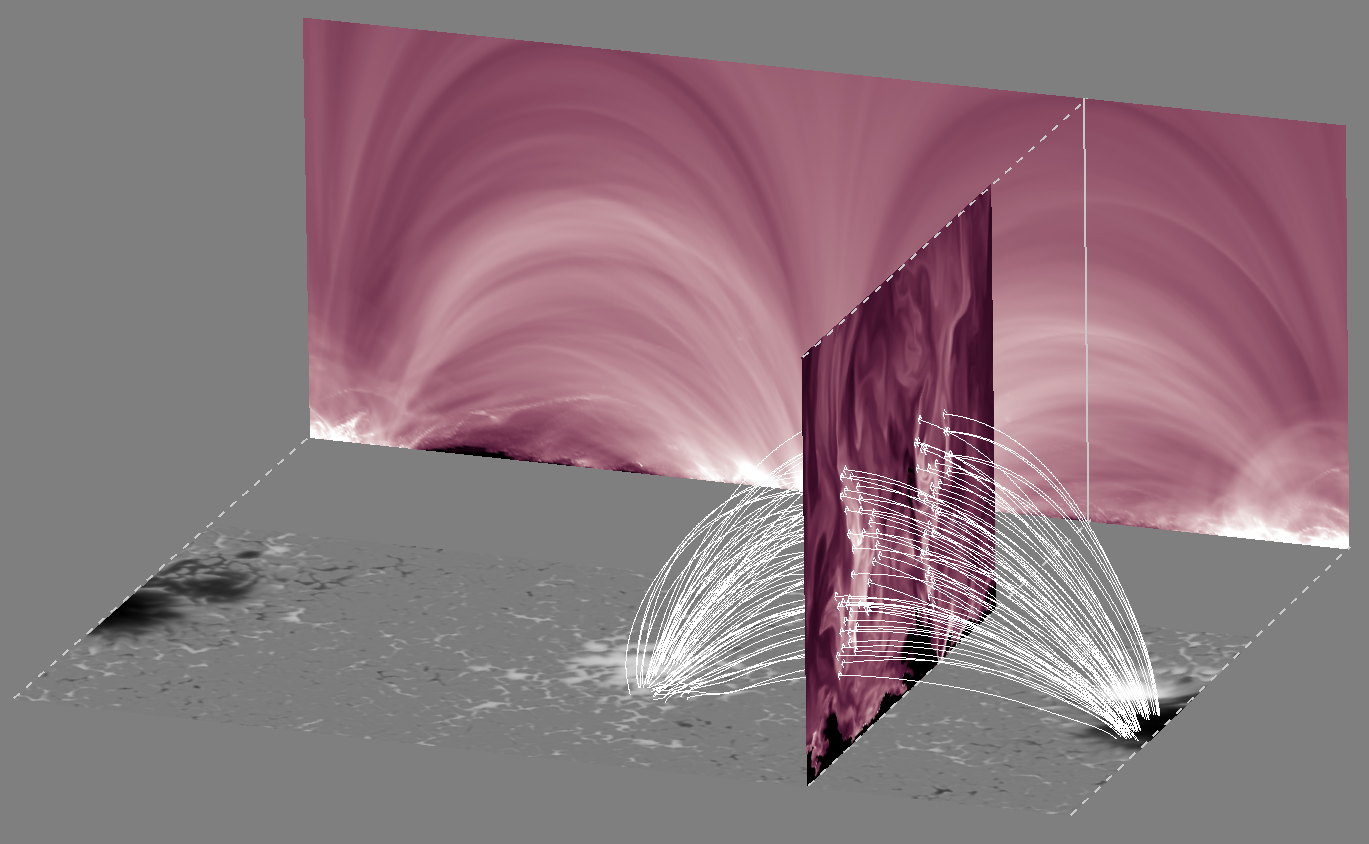}
\end{interactive}
  \end{center} 
 \caption{Magnetic field lines traced in the volume around several bright features in the slice from \figref{synth_midslices_x_384}, in the 211\AA~channel. The starting points are shown in \figref{bundle_x_384_locus}, left panel. An animated version of this figure is available in online materials. \nv{}{The animation shows the slice at different positions in the volume (left panel in the animation) as well as (right panel in the animation) the emissivity in the corresponding slice like in \figref{bundle_x_384_locus}.}}
 \label{bundle_x_384_3d}
\end{figure}

To address this question, we trace magnetic field lines around several bright features shown in \figref{synth_midslices_x_384}, in the 211\AA~channel. These field lines are shown in \figref{bundle_x_384_3d}. We then follow these field lines and examine the volumetric emissivity in their vicinity. The experiment is set up as follows. First, we select a $(y, z)$ vertical slice of $\vi$, chosen at the apices of a loop bundle, i.e., $x_0=24.6\mbox{ Mm}$, as in \figref{synth_midslices_x_384}. In this plane, we draw contours of $\vi(x_0, y, z)$ around the major bright features. These contours are, consequently, a set of points with coordinates $(x_0, y_i, z_i)$. From these points \textit{in the volume}, we initiate magnetic field line integration, forward and backward along the field. (For context, in many studies magnetic field lines are initiated from the photospheric boundary, e.g., from a set of points $(x_i, y_i, z=0)$. We instead use the points in the vertical plane, but the integration procedure is the same.) \p

\begin{figure}[h]
  \begin{center} 
  \includegraphics[height=5.75cm]{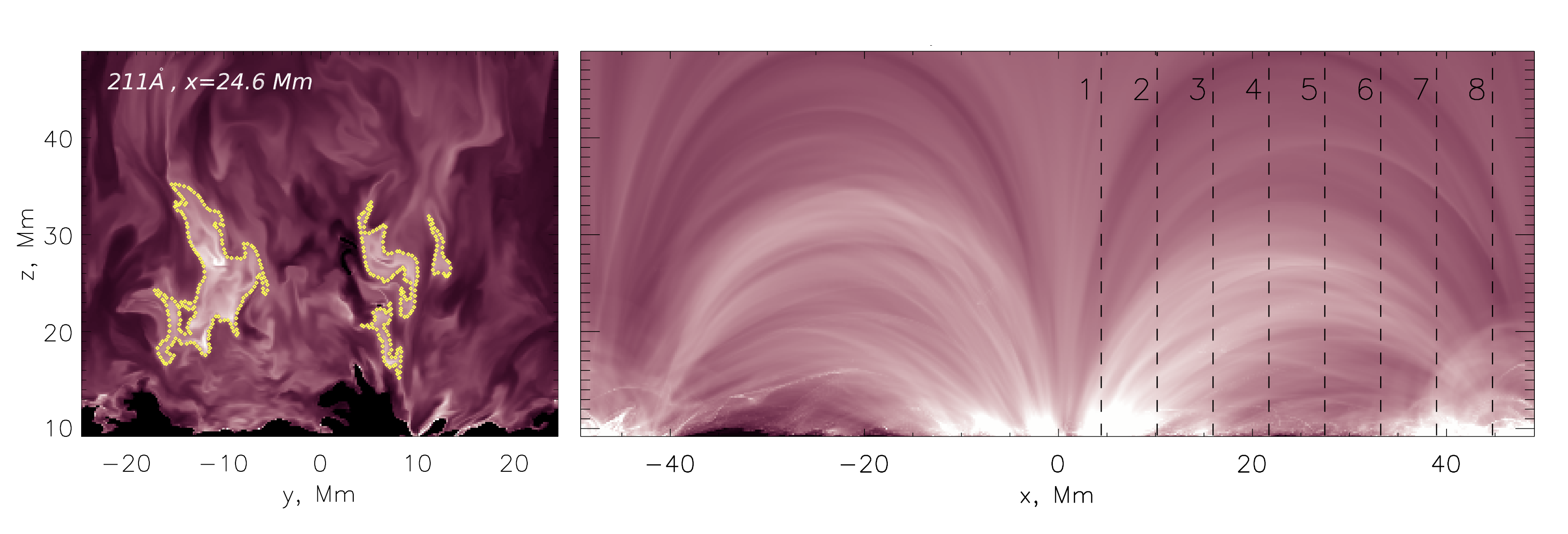}
  \end{center} 
 \caption{\textit{(Left panel)} -- volumetric emission at $\vi(y, z)$ at $x=24.6\mbox{ Mm}$, like in \figref{synth_midslices_x_384}. Dots mark the \textit{starting points} of magnetic field lines which are shown in \figref{bundle_x_384_3d}. \textit{(Right panel)} -- slices in the volume the field lines are then traced \textit{to}.} 
 \label{bundle_x_384_locus}
\end{figure}

\begin{figure}[h]
  \begin{center} 
   \includegraphics[width=17cm]{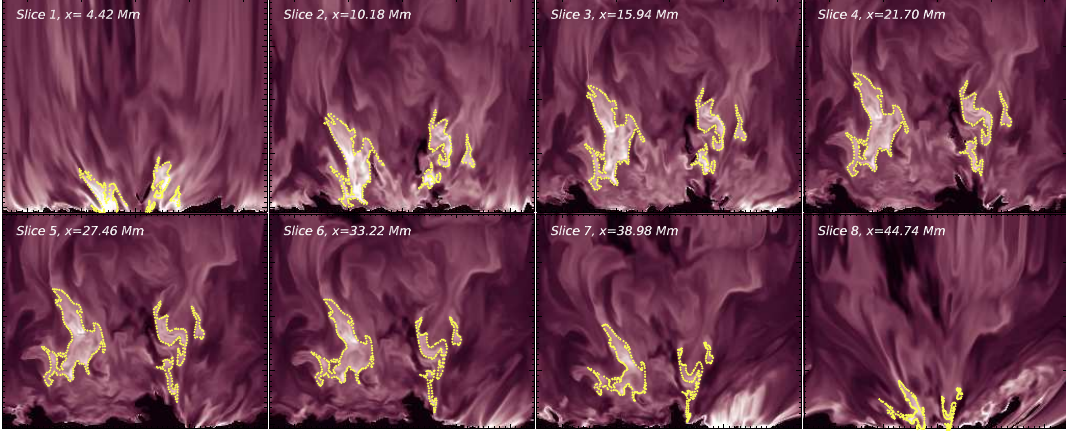}  
  \end{center} 
 \caption{Volumetric emissivity in vertical slices shown in the right panel of \figref{bundle_x_384_locus}. The bright features remain coherent through the length of the bundle. The dots indicate where selected field lines cross a particular slice. The starting points of these field lines are shown in the left panel of \figref{bundle_x_384_locus}. It appears that bright features enveloped by the field lines at one location, remain within the same field lines elsewhere in the volume. This means that the bright emitting features are indeed embedded in the flux tubes we construct; however, the cross-sections of these flux tubes are large (compared to thickness of strands seen in the synthetic line-of-sigh integrated images) and are of complex shape with wrinkled boundary.} 
 \label{synth_slices_x_384}
\end{figure}

The points on the contours are shown in the left panel of \figref{bundle_x_384_locus}. We then examine vertical slices of $\vi(x_i, y, z)$; their location within the domain is shown in the right panel of \figref{bundle_x_384_locus}. The slices themselves are shown in \figref{synth_slices_x_384}. In each slice, we mark the locations where the traced field lines cross the plane of the slice. \p

Two results appear immediately. First, the bright features stay coherent through consecutive slices. Second, \textit{the bright features appear to be enclosed by the same magnetic field lines on consecutive slices}. This is consistent with the picture of enhanced emission along flux tubes. However, the flux tubes containing emitting features, are: (1) large in diameter (compared to thickness of strands in synthetic AIA images); (2) fewer in number, compared to the amount of strands seen in the synthetic images; and (3) highly structured, with complex shape and a wrinkled boundary. Only the brightest of emission features, and consequently, the enveloping flux tubes, appear well defined; dimmer features appear spatially extended and difficult to separate from one another. The dim diffuse emission appears highly structured as well. \p

To summarize, the emitting structures are of complex shape, but they are embedded in magnetic flux bundles much as in \figref{synth_slices_x_384}. We therefore refer to these flux bundles as ``magnetic structures'' rather than ``flux tubes'', as the latter implies that the structures are (a) isolated, (b) small in cross-section compared to footpoint-to-footpoint length, and (c) simple in cross-sectional shape. None of these applies in general to the structures of continuous enhanced emission in the simulation we examine in this paper. Our observations, however, do not appear specific to the particular MURaM simulation: further in \secref{sec_conclusions} we point at independent simulations in which similar structures have been reported. \p 

\section{The Minimal-Isle Experiment}\label{sec_minimal_isle}

While the emitting structures in the volume do have complex structure, the brightest features (e.g., in Figures~\ref{synth_midslices_x_128} and~\ref{synth_midslices_x_384}) are indeed compact. Are they alone sufficient to explain the general appearance of coronal loops? We rephrase this question as follows: how much of the dim emission can be removed from the volume, so that the remaining islands of bright plasma maintain the pretense of the same loop structures, when integrated along the line of sight? \p

To determine this, we compute the magnetic mapping from the ``central'' Slices 1 and 2 to their photospheric footpoints. We then construct magnetic flux tubes which embed individual bright features on the central slices, for a choice of a brightness threshold and integrate the emission from the corresponding flux tubes only, masking out the rest of the volume. Finally, we vary the threshold and compare the resulting synthetic observables with those of the full simulation volume. The maximal emission threshold (and consequently, smaller islands of emission) for which the same loop structures can be observed (as in the full cube rendering) would \nv{signify}{define} the minimal islands of emission significant for the overall observed configuration of loops.\p

\begin{figure}[h]
  \begin{center} 
   \includegraphics[height=7cm]{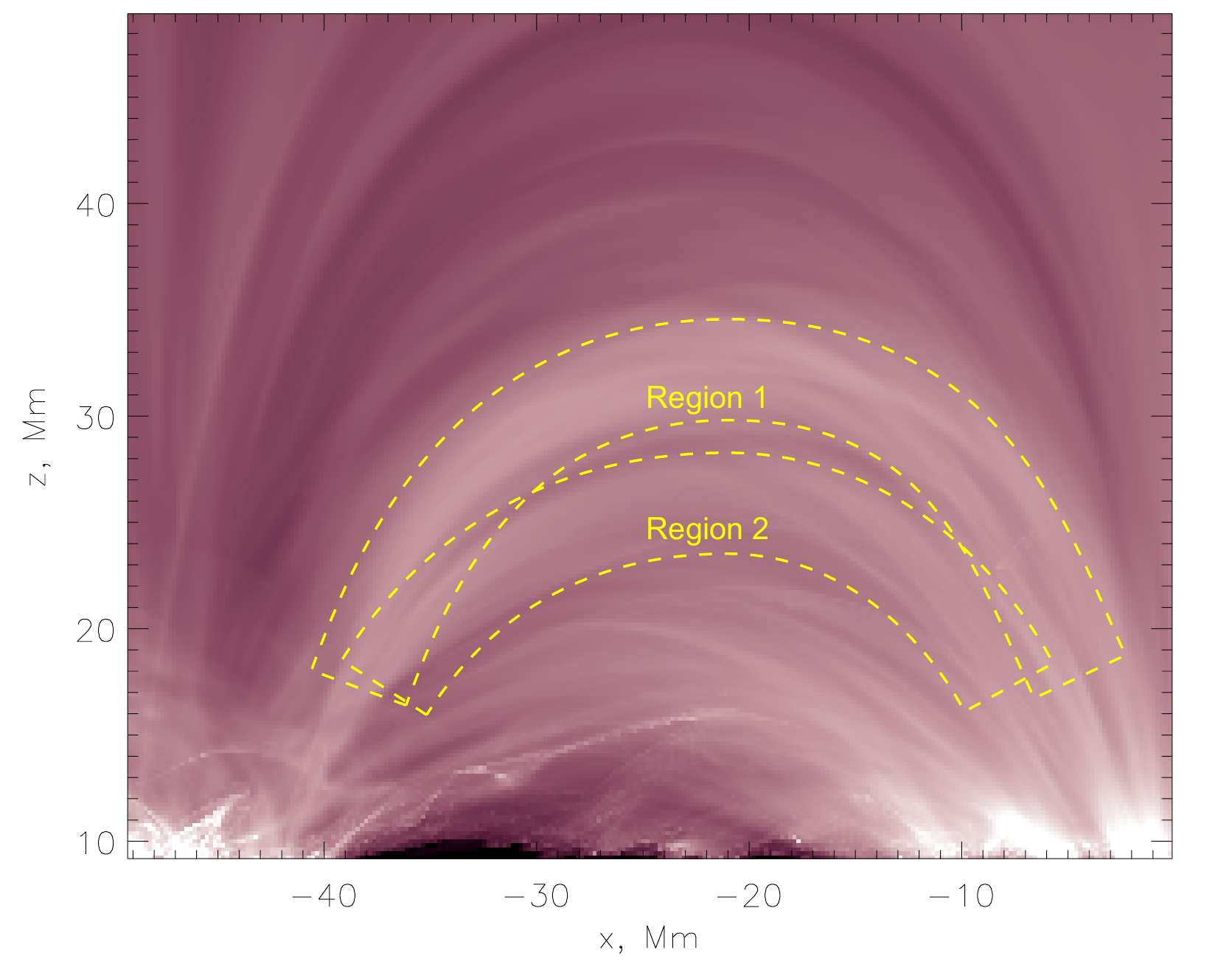} 
  \end{center} 
 \caption{Two curvilinear regions surrounding bundles of loops which we focus on in \secref{sec_minimal_isle}. (Note that as the bundles converge, there is a small overlap between the two regions.)}
 \label{fig_slab_regions}
\end{figure}

\begin{figure}[h]
  \begin{center} 
   \includegraphics[height=7.5cm]{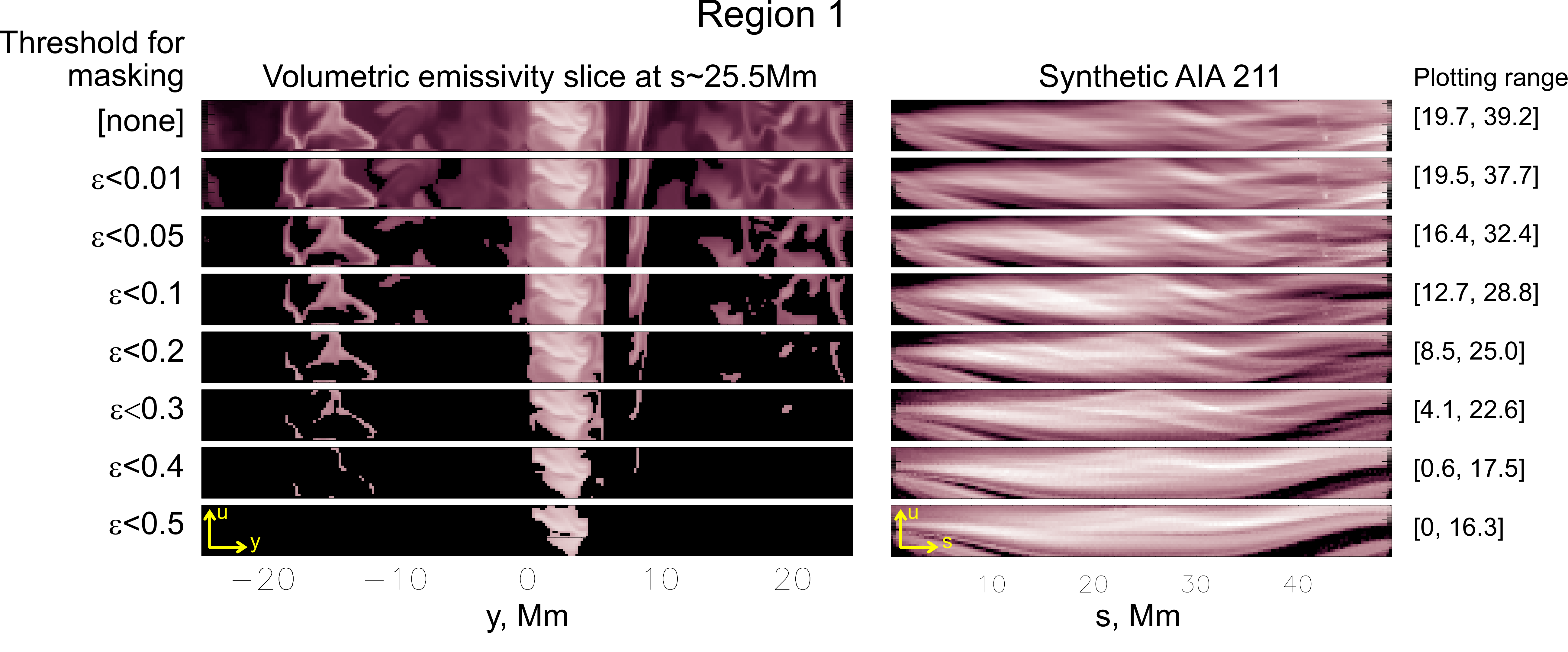} 
  \end{center} 
 \caption ~Region 1 from \figref{fig_slab_regions}. \nv{Left}{Right} column: synthetic 211\AA{} emission $I(s, u)$, straightened out along the long edge ($s$). The horizontal axis is along $\shat$, and vertical axis is along the direction $\uhat$, locally normal to $\shat$. \nv{Right}{Left} column: volumetric emissivity $\epsilon(y, u)$ on a plane along the line of sight (i.e., perpendicular to the images in the \nv{left}{right} column) at $s\approx 25.5\mbox{ Mm}$, to illustrate the kind of features which remain, as dim emission is masked out of the volume at progressively higher thresholds. The images in the \nv{left}{right} panels show AIA 211 channel synthesized after the dim emission has been progressively masked out. Note how the observed morphology of the loops changes with the removal of the dim emission; the image in the last row is a subset of the image in the first row, but it does not seem to be responsible for a large number of bright features in the original loop bundle. \nv{}{Note that synthetic images of $I(s,u)$ are plotted using a progressively narrower intensity range, $[\mbox{min}, \mbox{max}]$, as indicated in the right hand column. This is done to ensure remaining features are visible.}\footnote{\nv{}{Both $I$ and $\epsilon$ are plotted on the square-root intensity scale, although the line-of-sight integration is done on a linear scale. The units of $\epsilon$ are arbitrary, and the images of $\epsilon(y,u)$ in the left hand column are plotted in the range of $[\mbox{min}=0.0, \mbox{max}=1.2]$ in these units. The images of $I(s,u)$ are in the same units times length in pixels; the volume is 256 pixels in the $\yhat$ direction.}}
 \label{fig_slab_1}
\end{figure}

\begin{figure}[h]
  \begin{center} 
   \includegraphics[height=7.5cm]{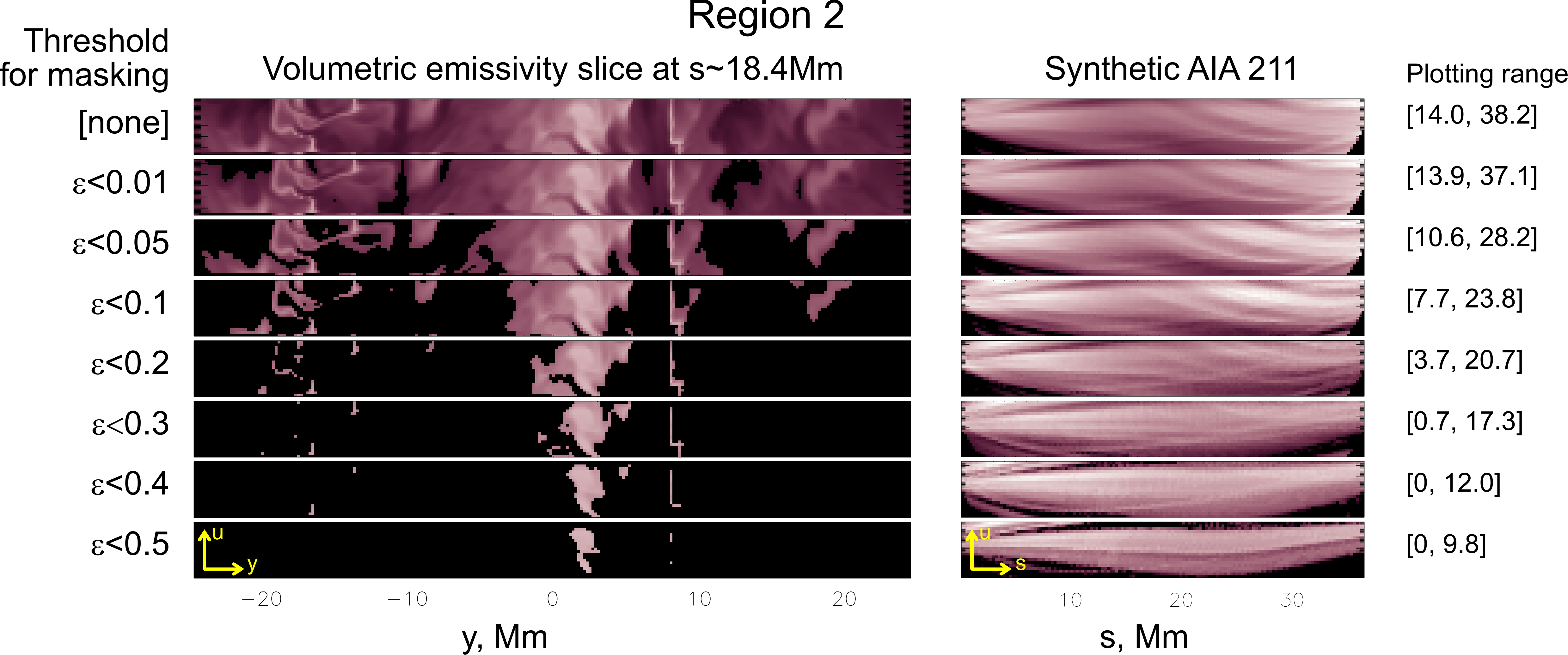} 
  \end{center} 
 \caption{Region 2 from \figref{fig_slab_regions}, the notation is the same as in \figref{fig_slab_1}. (The volume slice here is taken at $s\approx 18.4\mbox{ Mm}$.)}
 \label{fig_slab_2}
\end{figure}

\figref{fig_slab_regions} shows two regions that we focus on. Figures~\ref{fig_slab_1} and~\ref{fig_slab_2} show these two regions ``straightened up'' (in coordinates $(\shat, \uhat)$, $\shat$ being the direction along the loop bundle, and $\uhat$ being locally normal to $\shat$). These images are recalculated with dim plasma in the volume masked out, while progressively increasing the threshold for such masking. In both of these figures, the \nv{second}{left hand} column corresponds to the $(y-u)$ volumetric emissivity slice through the middle of the region. As is evident from these figures, the emission from the bright compact features is not responsible for the impression of many bright strands seen in the unmasked images. The particular parts of the volume that correspond to a particular strand in the image are examined in more detail in the next section and in the appendix, while here, we would like to draw attention to how different is the appearance of the loop bundles with only compact emission taken into consideration. While the strands which are the result of these small compact features are present in the ``unmasked'' image, they are not the most noticeable, nor are they the most common in this particular observation. The overall shapes, lengths, and directions of loops in the ``unmasked'' image seem to be largely determined by the dim features of complex shape.\p 

\section{The Wrinkles in the Veil}\label{sec_single_loop}

As we demonstrated in the previous sections, the volumetric emission has a complicated structure. Emission in Figures~\ref{synth_midslices_x_128} and~\ref{synth_midslices_x_384} is structured into shapes akin to puffs of smoke. Many bright substructures appear in the form of thin ragged and wrinkled sheets, akin to those of a \textit{``veil curtain''}. \p 

\citet{Malanushenko2013} have previously pointed out that structures which are elongated along the LOS will appear thinner \textit{and} brighter than those elongated across the LOS. Consequently, a curtain-like structure which is thin and wrinkled may produce \textit{several} features when integrated along LOS: thin and bright shape(s) where the wrinkles locally coalign with the LOS, superposed over dim diffuse background where the structure is locally perpendicular to the LOS. \p

\begin{figure}[h]
  \begin{center} 
   \includegraphics[width=10cm]{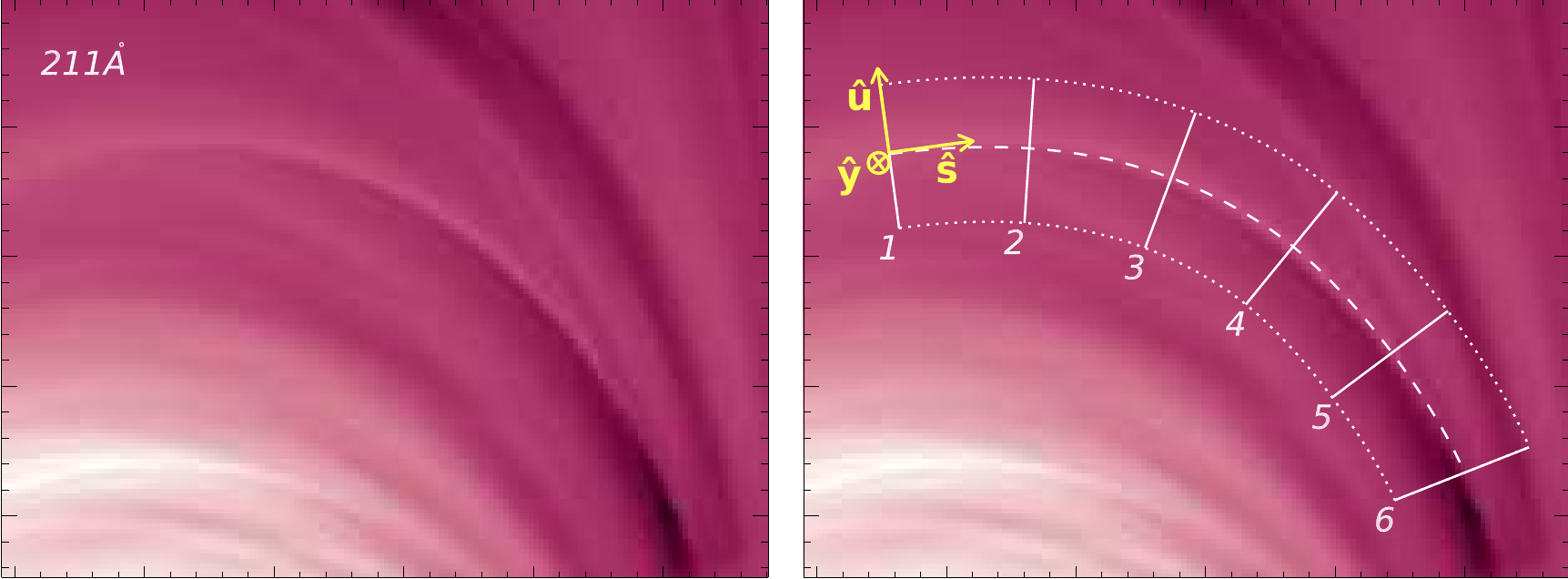}  
   \end{center} 
 \caption{\textit{Left panel:} subregion in the synthetic 211\AA~image which shows a relatively clear and isolated loop. \textit{Right panel:} the traced loop, the slab around it, the slices across the loop which we will examine, and the coordinates which will be further used.}
 \label{loop_pos}
\end{figure}

We illustrate this effect on a particular loop seen in the 211\AA~channel of the synthetic AIA images. The loop is shown in \figref{loop_pos}. We attempt to locate the bright structure in the volume which corresponds to this loop. For that, we first manually draw a smooth curve along the loop in the plane-of-sky (POS). \nv{}{For this purpose, we adopt the definition of a loop as the smallest quasilinear visual feature in the corona, as has been and remains common practice in relevant studies \citep[e.g.,][etc]{Testa2002, Aschwanden2005, Aschwanden2017, Kucera2019, Klimchuk2020, McCarthy2021}.} We then define a thick slab surrounding this curve, in the same manner as in the previous section, and extract a 3D subvolume which projects into this slab when integrated along the LOS. We then analyze slices in this subvolume. \p

The set up of this experiment is as follows. Let the direction along the curve be $\shat$ and let the LOS direction be $\yhat$; the coordinate in the POS across the loop, $\uhat$, completes the orthogonal basis. These coordinates are labeled in \figref{loop_pos}. We choose the following ranges: $s\in[0, s_{max}]$ is where the loop is clearly visible, $u\in[-u_{max}/2, u_{max}/2]$ as shown in \figref{loop_pos}, and $y\in[y_{min}, y_{max}]$ being the full extend of the volume in the $\yhat$ direction. We then extract and straighten the curvilinear 3D slab, akin to ``straightening the loop'' \citep{DeForest2007, Winebarger2014}, but with the inclusion of the third dimension. \p

\begin{figure}[h]
  \begin{center} 
   \includegraphics[width=17cm]{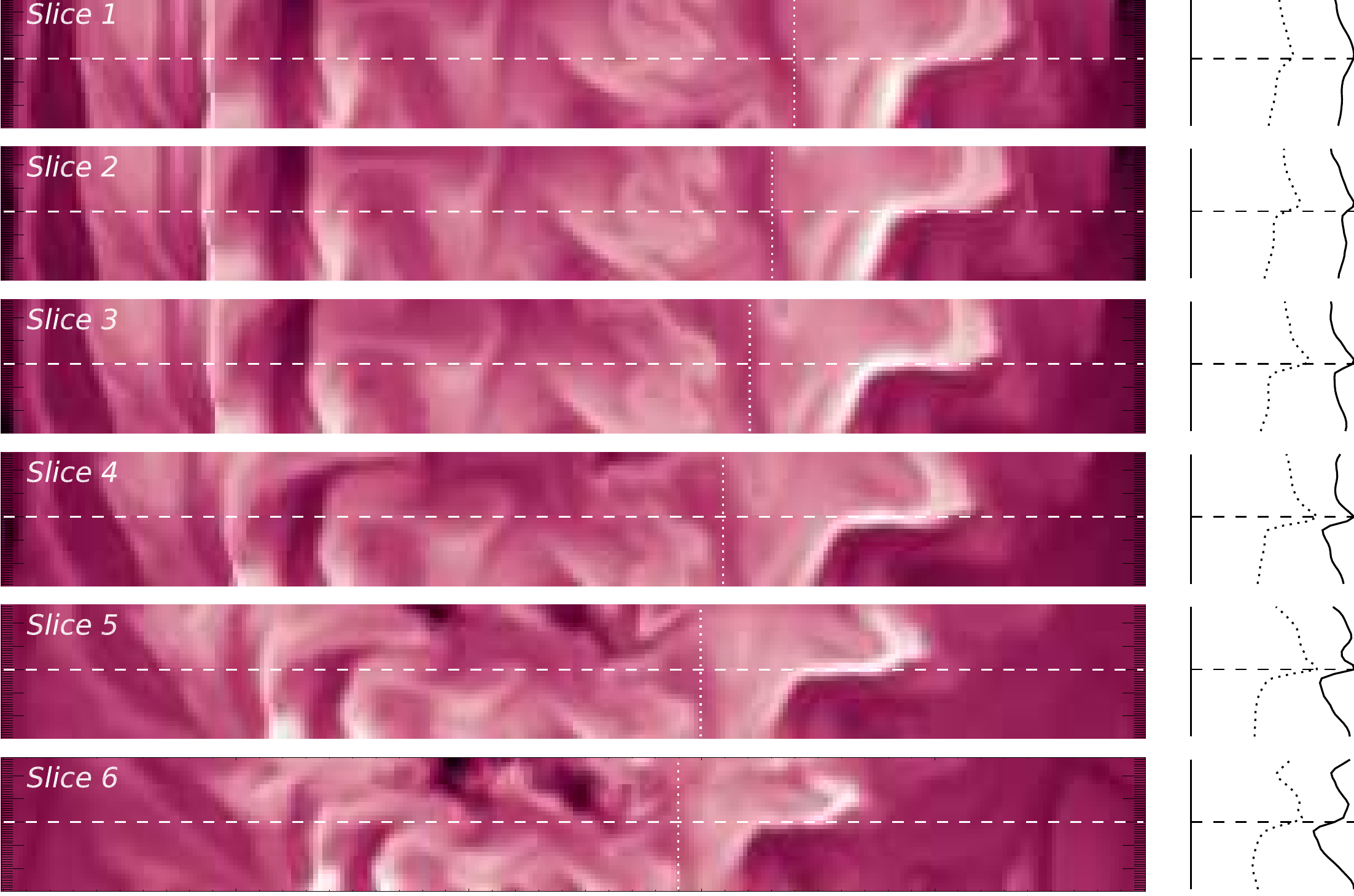}  
   \end{center} 
 \caption{Slices in the volume which project into the lines 1-6 shown in \figref{loop_pos}. The right panels here show, for each respective slice, it's integral over $\yhat$, which is equivalent to the values of intensity along the lines 1-6 in \figref{loop_pos}. The solid line shows the integral over the entire $y$ range, while the dotted line shows the integral of the subregion \textit{to the right} of the vertical dotted line shown in each slice \nv{}{(both curves are plotted with respect to zero, marked as solid vertical line, to give insight on the relative contrast of the bump)}. This subregion, in each slice, is meant to isolate the \zz-shaped feature which we argue is what contributes the most to the loop from \figref{loop_pos}.} 
 \label{loop_slices}
\end{figure}

\figref{loop_slices} shows $(y, u)$ slices of $\vi$. For each slice, its LOS integral is shown on the side to guide the reader. The latter is the same as intensity profile in the LOS-integrated image along lines labeled in \figref{loop_pos}. Firstly, we notice that in respective slices, the features stay coherent, with small changes in shape, size, and position from slice to slice. Secondly, we notice a particular sheet-like feature with a bend in it: a \zz-shaped\footnote{The \zzs is the archaic Greek letter ``Koppa''.} feature in the right part of the slices (approximately in $y\in[30, 45]\mbox{ Mm}$). On each slice, we isolate this feature by computing the $\yhat$-LOS-integral over the corresponding subregion: from $y_{min}$, which is shown as a dotted vertical line in each slice in \figref{loop_slices}, to $y_{max}$, which is the edge of the entire volume. For each slice, the LOS-integral over the subregion $y\in[y_{min}, y_{max}]$ is plotted in the right panel in dotted line.\p 

By looking at the slices in \figref{loop_slices} and at both LOS-integrals to the right of each slice, we notice the following. The dashed and the solid lines in the right panels are close in shape, which suggests that \zzs feature is indeed what is responsible for creating the loop we study. \nv{}{The loop itself appears to be a small enhancement over the background, as has been previously found in actually observed loops in the literature \citep[e.g.,][]{Klimchuk2020}.} We also notice that the \zzs distorts slightly from slice to slice. It seems that its horizontal portion is aligned with the LOS ($\yhat$, the dashed line in the images in left panels) best between slices 2 and 5. It is noteworthy that the loop itself, shown in \figref{loop_pos}, appears to be the most well-defined between slices 2 and 5 as well. \p

All this analysis leads us to the conclusion that the loop we study is in fact a projection artifact, a \textit{``wrinkle in the veil''}. One might argue that the horizontal portion of the \zzs could be considered a stand-alone loop. However, as the entire feature rotates, and the thickness of the projected loop changes, other parts of the \zzs begin to stand out of the background. \nv{}{Note that the particular portion of \zzs which manifests as a loop is actually line of sight dependent. Moreover, a non-zero magnetic twist would exacerbate the situation: if a \zz-shaped flux tube is twisted, then along the flux tube's length, \textit{different} parts of the cross-section would manifest as a strand. To an unsuspecting observer a large amount of twist would even manifest as one strand disappearing midway and two more appearing slightly offset.} Additionally, it appears that the distribution of the brightness along the feature changes: in Slice 1, it's lower vertical portion is the brightest; by Slice 5, the the middle horizontal portion becomes the brightest, and in Slice 6, its upper vertical portion is the brightest. This may be the result of the effects described in \citet{Peter2012}, slight temperature/density variations in different portions of the flux tube. \nv{Regardless of whether this effect is responsible for the brightness variation,}{For these reasons,} we are inclined to consider the entirety of this feature as one ``flux tube'', as it spins around, gets distorted and changes in size \textit{as a whole}. \p 

\section{Implications for Stereoscopy}\label{sec_stereo}

While it does \textit{not} appear that \textit{all} structures in our simulation are thin wrinkled sheets, \textit{many} of them are. As we showed above, the wrinkles can produce the impression of compact loops, and it is not clear how to tell them apart from compact blobs based on the line-of-sight integrated observations alone. This means that for any observed loop, the possibility of it being a projection of a wrinkle must individually be ruled out.\p 

\begin{figure}[h]
  \begin{center} 
   \includegraphics[width=12cm]{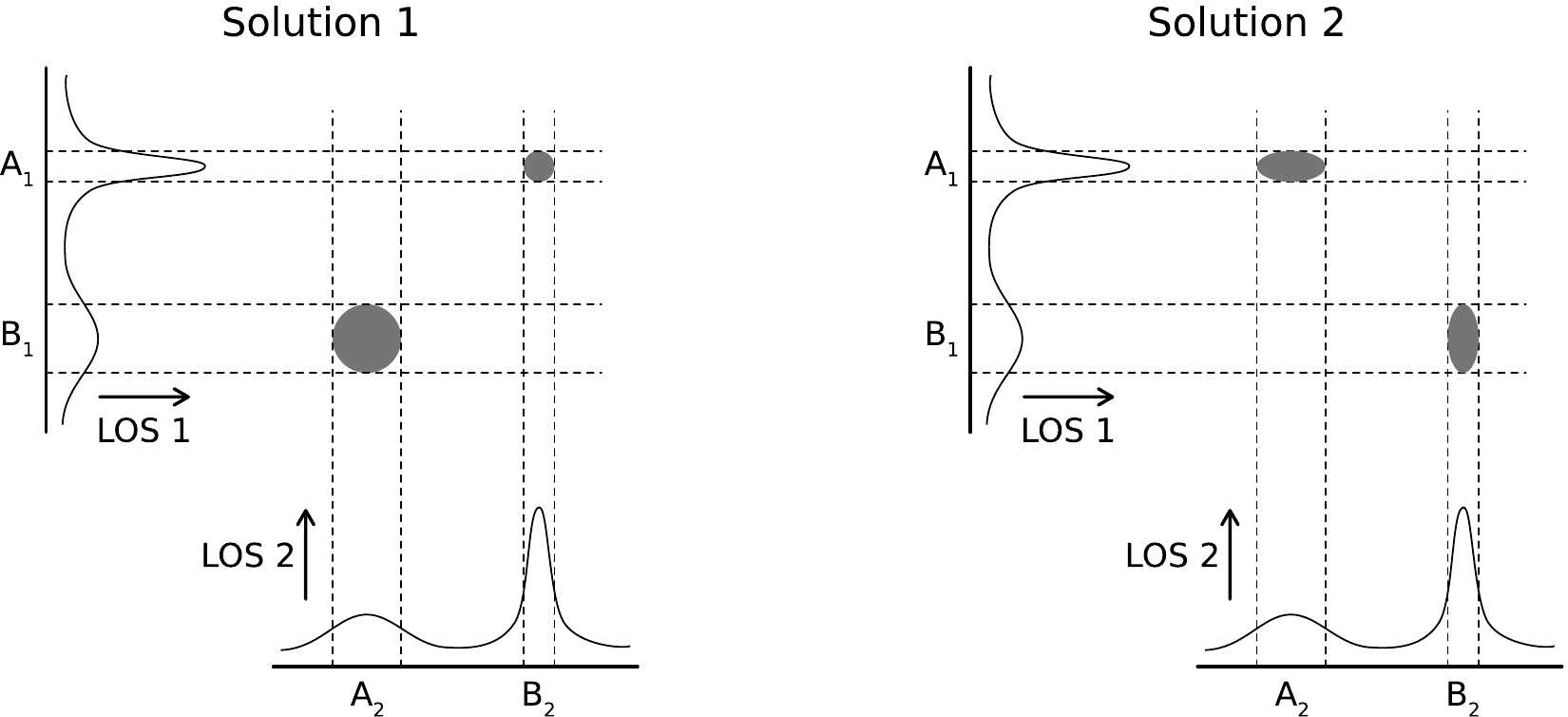}  
   \end{center} 
 \caption{A brief recap of the argument made by \citet{Malanushenko2013} with regard to the triangulation of compact structures, but when the cross-section is not known to be circular \textit{a priori}. Suppose the same two structures are observed by two instruments with lines-of-sight $90^\circ$ apart. Suppose each sees a thin bright loop and a thick dim loop, in a similar location and with similar trajectories. (We refer the readers to the Figure~11 of that paper for a rendering of such a pair of loops.) When doing triangulation, there would be two different solutions (left and right panels), which will produce exactly the same observations in both imagers; it is therefore not possible to tell which one is valid from these two views alone.}
 \label{twoloops_stereo}
\end{figure}

While it may appear at first sight that stereoscopy may help telling apart compact flux tubes and projections of wrinkles, we believe that it may instead produce additional difficulties. \citet{Malanushenko2013} have argued that even if structures were known to be compact, they may have elongated cross-sections, which in turn may complicate stereoscopical determination of their shapes. Their argument went as follows. Suppose there are two instruments imaging the same region from two different view points, such as the case for the twin STEREO satellites \citep{StereoRef}, suppose for simplicity of argument the lines of sight are $90^\circ$ from one another, and suppose each instrument sees two nearby loops, A and B. In order to proceed with triangulation, the observer must first decide which of the loops in one imager corresponds to which of the loops in the other imager. If the trajectories of A and B \nv{}{and their total flux integrated over the cross-section} are similar, then there are in principle two options, both of which produce acceptable trajectories: $(A_1\rightarrow A_2, B_1\rightarrow B_2)$, $(A_1\rightarrow B_2, B_1\rightarrow A_2)$. It may seem reasonable to make the choice based on other properties, such as loops' diameters and brightnesses. \citet{Malanushenko2013} have argued that this is in fact can be \textit{misleading}, \textit{if} both flux tubes have elongated cross-sections. The two solutions are shown in \figref{twoloops_stereo}. It appears that without an \textit{a priori} knowledge of whether the structures are elongated or not in cross-section, it is not possible to tell these two solutions apart based on the two views alone. The addition of the third viewpoint may in principle help; however, there are usually many more than two loops in an active region, which may lead to more than just two degenerate solutions. \nv{}{In the idealized setup, the ambiguity could be broken by different total flux in the two structures (i.e., total area under $A_1$, $A_2$, etc). However, in real observations there will generally be background (generally nonuniform), noise, etc. Therefore when an observer is tasked with finding matching structures in multiple imagers in real data, the allowance for the observational uncertainties is typically made \citep[e.g.,][]{Kucera2019,McCarthy2021}}.\p

\begin{figure}[h]
  \begin{center} 
   \includegraphics[width=5cm]{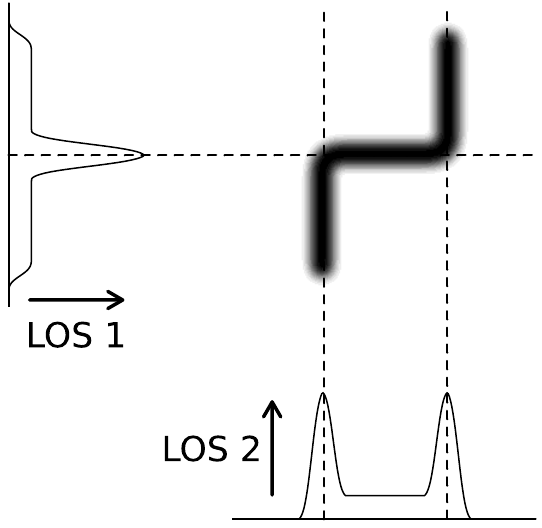}  
   \end{center} 
 \caption{Implications of existence of bright wrinkled sheets in the corona for stereoscopy. Here a conceptual rendering of the \zz~feature from \figref{loop_slices} is integrated along two different lines of sight (as if observed by the STEREO spacecrafts when they are $90^\circ$ apart). An observer looking along LOS 1 will see \textit{one} loop and background \textit{around}. An observer looking along LOS 2 will see \textit{two} loops and background \textit{in between}. The ground truth, however, is a single structure, and it contributes to both the loops and the background; not only will the observers see different amount of features, but also subtracting the background will subtract a portion of the structure itself.}
 \label{wrinkle_stereo}
\end{figure}

We now turn to the \zz-shaped structure discussed above and shown in \figref{loop_slices}. Under a similar observational setup, shown in \figref{wrinkle_stereo}, one can follow a similar line of arguments. However, in this case the \textit{number} of observed loops will differ: the first imager will see one loop (surrounded by background), and the second imager will see two loops (with background between them). A second solution of the triangulation problem given this pair of observations may be that there are two loops, and that they are coaligned along LOS1. However, if that were the case, then the total intensity of the peaks that the second observer sees should be equal to the intensity of the peak that the first observer sees. It is evident from the figure that this will not in general uphold for a wrinkled structure, because the two imagers will see \textit{different parts} of the structure as peaks. This will be especially true if background subtraction is used: what one observer sees as a peak, the second sees as a background, and \textit{vice versa}. \p

The addition of a third view point can in principle help resolve the problems we here outline. But it is likely that additional studies will be needed to confirm or reject this possibility. \p 

\section{Summary of the results}\label{sec_results}
We have studied a 3D MHD simulation of an active region created with the MURaM code. We focused on the coronal portion of the datacube and studied synthetic EUV observation and volumetric emission of plasma. Our findings are as follows. 
\begin{enumerate}
	\item{The synthetic observations in six AIA EUV channels appear realistic. We focused on the two major loop bundles. Each of these bundles appears to be composed of many individual loops. These loops have various thicknesses, the thinnest of them being near the resolution of the simulation itself. Most of the loops appear as incomplete arches. Many loops are but a small enhancement of the background intensity, which is consistent with observations \citep[e.g.,][]{Klimchuk2020}. The loop tops appear bright and thin, which is \nv{}{qualitatively} consistent with observations of non-expanding loops with high pressure scale height.}
	\item{We then studied emitting structures in the volume (which need to be integrated along the line of sight to produce the synthetic observables). The expectation, based on traditional picture of individual loops being the projections of individual thin flux tubes, was to see many thin and isolated structures with compact cross-sections. The structures in the volume that we observe are at a striking disagreement with this expectation. Overall we observe fewer structures of larger size. These structures have remarkably complicated cross-sectional shape and complex, ragged boundaries. The complexity of the shapes is such that in many instances, it is difficult to separate different features, as they appear to blend and merge into one another. Some of the bright substructures appear to be in the shape of thin and long sheets, ragged and wrinkled, also merging into one another.}
	\item{We were able to confirm that these emitting structures nonetheless follow lines of magnetic field. We did this by drawing contours around some features at some location in the volume, initiating magnetic field lines from the points along these contours, and following both the field lines and the emission through the volume. This result means that the complicated emitting structures we observe \textit{are}, in fact, magnetic flux tubes; however, neither are they \textit{thin}, nor their cross-section is \textit{compact} and well-defined compared to their surroundings.}
	\item{We studied one particular loop which appeared to be reasonably well visible and reasonably isolated from other loops. We were able to identify a single emitting structure in the volume which mapped into this loop. Its cross-sectional shape was that of a thin sheet with two bends in it, akin to the Greek letter Koppa (\zz). Its middle portion was oriented along the line of sight and we found that this middle portion was well correlated with the location of the loop in the LOS-integrated image. The entire feature was rotating along the length of the loop, and the loop was best defined when the middle portion was co-aligned with the line of sight. This implies that the loop we observed was not a projection of a thin, compact flux tube, but rather it was a line-of-sight artifact, a projection of a wrinkle of a veil-like structure.}
	\item{We argued that some unknown number of loops which appear in these synthetic images may in fact be such artefacts: projections of emitting veils in the volume, rather than thin isolated flux tubes. It is not clear how to tell one from another given only the line-of-sight integrated images. Appendix A shows analysis similar to that of the \zzs structure, but for a number of coronal loops, suggesting that loops that are projection artifacts are ubiquitous.}
	\item{We also demonstrated that stereoscopy will not necessarily help rule one over the other in each particular case. Instead, wrinkled shapes may introduce additional problems in triangulation procedure. These problems will not necessarily manifest themselves in any noticeable way, short of yielding results which do not correspond to the ground truth.}
\end{enumerate}

\section{Can We Believe \mm{}?}\label{sec_other_mhd}
The \mm{} simulation we studied here is one of the most realistic coronal simulations available to date. However, it is certainly prone to criticism, as any other simulation. We nonetheless believe that our results are relevant to the actual corona, for the following reasons: 
\begin{itemize}
	\item{The simulation captures much of the known physics, and successfully reproduces many observational phenomena (see \secref{sec_muram}). The combination of these two factors is in principle enough to consider the model plausible. This is in fact is a typical line of arguments in many studies which are focused on simulations.}
	\item{Additionally, we notice that MURaM is far from the only simulation in which the ``veil curtains'' of emission have been noted. Below is the list of \textit{independent} simulations, which feature independent realizations of \textit{both} the \textit{ab initio} full-MHD, and also other kinds of models, in which comparable features have been reported. They were observed in various scales and resolutions, from ultra-fine resolution models several Mm across, to full-AR size solutions: 	
	\begin{itemize}
		\item{\citet{Gudiksen2005b} used their 3D MHD model to simulate a corona above a small active region. The velocity field at the lower boundary was designed to mimic granulation pattern, and the heating was ohmic, induced by small flow motions at the lower boundary. The resulting synthetic images are generally in good agreement with observations. Most of the features discussed in their study are in the plane of sky, but the authors do briefly touch the topic of cross-sectional properties of coronal loops. They find that for a randomly selected set of flux tubes which are circular at the apices, the footpoints are strongly distorted \citep[which agrees with][]{Malanushenko2013}. \p 
		Most importantly for the coronal veil hypothesis, \citet{Gudiksen2005b} say the following: ``In general the footprints of the loops shown in Figure 3 are wrinkled and not geometrically simple, and their area cannot be approximated by product of $\Delta X$ and $\Delta Y$. We expect the footprints to increase in complexity with increasing resolution, which may give the \textit{impression} that the cross-sectional area changes less than it actually does.''.} 
		\item{\citet{Mok2005} have developed a unique method for simulating coronal emission, in which the magnetic and the plasma processes are intrinsically decoupled. First, a model of a magnetic field is calculated using a $\beta=0$ assumption. Then, for every voxel in the volume, a magnetic field line is traced and hydrodynamic equations are solved along this field line. While this models makes more approximations than full-MHD models make, its advantages include the fact that the cross-field diffusion and conduction are null by design (which is difficult to achieve in a standard 3D simulation). Additionally, it allows the user to prescribe different heating functions corresponding to different analytical coronal heating models. \p
		In a recent use of this model, \citet{Winebarger2014} simulated coronal emission based on a magnetogram from a particular active region and used a quasi-steady highly-stratified heating function which, as they demonstrate in previous papers, produces a good match to observations. They identified and analyzed several loops on the synthetic coronal images. They were able to identify the distinct structures in the volume which produced these loops. \p
		They also show a vertical slice of volumetric emission in their Figure~4 (similar to our Figures~\ref{synth_midslices_x_128} and~\ref{synth_midslices_x_384}). It demonstrates features qualitatively similar to those we show in the current paper: the emission is composed of (a) diffuse component, (b) several isolated blobs, and (c) thin, bright, sheet-like structures. \citet{Winebarger2014} focus specifically on loops which are due to flux structures of compact, round cross-section. The wrinkles in the sheet-like structures are not commented upon, but from visual examination of their Figures~1 and~4 it appears that there are in fact loops that are mere projection artifacts due to these wrinkles, much like our example in \secref{sec_single_loop}.} 
		\item{\citet{Antolin2014} studied evolution of a thin straight flux rope subject to transverse MHD waves. In their simulation, they had a thin cylindrical flux rope in hydrostatic equilibrium. The flux rope was then subject to a perturbation consistent with a kink mode. As the axis of the flux rope wriggled on a large scale, it turned out that the boundary of the flux rope, initially circular in cross-section, developed small current sheets and consequently fine structure around its circumference. \citet{Antolin2014} argued that this is a manifestation of Kelvin-Helmholtz instability, caused by sheering motions and exacerbated by resonant absorption. They also pointed out that the eddies generated by the Kelvin-Helmholtz instability may produce fine-scale strands in synthetic images, causing the impression of fine substructuring within the initial cylindrical ``loop''.\p
		Qualitatively, the boundary of their flux rope (their Figure~3) appears to have ragged and wrinkled edge, and is consistent with the ``veil'' hypothesis outlined in our paper. The cross-sectional shape of their flux rope is in general consistent with the smaller-scale structures in our Figures~\ref{synth_midslices_x_128} and~\ref{synth_midslices_x_384}. However, the effect they study takes place on extremely small scales: their flux rope is $\approx 1\mbox{Mm}$ in diameter (and the eddies are much smaller than that), while the size in cross-section of the flux structures we show in \figref{bundle_x_384_locus} are $\approx 10\mbox{Mm}$. Also, while Kelvin-Helmholtz instability plays a major role in the formation of the eddies in \citet{Antolin2014}, it is unclear whether this is a major effect in our study. It must be noted that while \citet{Antolin2014} \nv{evolves}{evolve} a flux rope initially smooth and round in cross-section, our simulation is \textit{ab initio} and contains no such pre-determined setup in the corona. The movies of the time history of flux bundles in our simulation suggest that they form with an \textit{already} complex cross-section, rather than evolve to it from an initially simple and smooth shape.} 
	\end{itemize}	
	We stress that to our knowledge, these codes were developed independently from one another. The only thing they have in common is the physics they approximate. Therefore, it is likely that it is the physics which leads to the formation of such features.}
\end{itemize}

\nv{Finally, we note that even if \mm, along with other MHD codes we mention, made predictions which were completely wrong (which we argue is unlikely), it really \textit{does not matter} for this study.}{Finally, we note that the core implication of this paper about veils having similar signatures as compact features, despite illustrated with \mm, \textit{does not depend on \mm{} having a valid corona}, for reasons explained below and further in \secref{sec_conclusions}.} We use these data to illustrate how a combination of complex temperature and density distributions and line-of-sight integration can lead to the appearance of loop like features in synthetic emission even when the 3D volume data \nv{does}{do} not contain compact density and temperature structures \nv{itself}{themselves}. The main findings of our analysis are independent from a particular simulation setup and simply illustrate the intrinsic complexity involved in interpreting observations that result from line-of-sight integration in an optically thin plasma.\p

\nv{}{On a separate note, \textit{if} \mm{} \textit{does} in fact reproduce the actual corona, then our results imply the abundance of veils in the actual coronal emission. This raises a number of questions regarding their nature and the mechanisms of their formation. More specific questions include, for example, the following. What determines their thickness? What determines their complexity? Are there preferred places of their formation? Most importantly, is there a way to tell, for a particular loop, whether it is a veil or a structure with compact cross-section? These questions must be addressed in future studies.}

\section{Conclusions and Discussion}\label{sec_conclusions}
This work forces us to rethink the way we interpret observations of the solar corona. We demonstrate how a realistic-looking set of coronal loops can be a product of volumetric structures which are not consistent with the standard paradigm of loops representing individual thin bright structures confined to thin magnetic flux tubes. For a particular active region on the Sun this interpretation might or might not apply (however, the success of the \textit{ab initio} simulation from \cite{Cheung2019} is encouraging in this regard). We do not offer a quantitative statistics of ``loops'' vs. ``veils'', as this must be a study of a larger scope. However, the complex manifolds we observed in the MHD simulation are also a more general class of structures than individual isolated thin flux tubes, in the same sense as, for example, polygons are a more general class of shapes than triangles. This adds additional plausibility to our hypothesis. \p

While the concept of emitting veils \nv{}{in application for coronal loops} is relatively unexplored to date, and the strength of the case we make in its support constitutes an important result on its own, the appreciation of this fact may be distracting from the main take-home message we wish to offer to the readers. The most impactful conclusion of this paper, for practical purposes, may be that just by looking at an individual loop structure, we cannot tell whether it \textit{is} a veil, or whether it is a individual, coherent, compact structure viewed over a diffuse background. The visual appearance can be deceiving, as we demonstrate in \secref{sec_single_loop} and in Appendix~\ref{sec_appendix}. The apparently sane stereoscopic measurements may be deceiving as well, as we argue in \secref{sec_stereo}. The reason why this is important is that many measurements of the physical conditions in the solar corona, such as pressure scale height, temperature, measurements inferred from observed loop oscillations, and so on, make a tacit assumption that a given loop is ``real''; should the loop in question turn out to be a ``wrinkle in the veil'', the results, while appearing sensible, may be misleading.\p

Another consequence of ``the veil'' hypothesis regards background subtraction. Even for an isolated loop, subtracting what \textit{appears} to be background will in fact result in removing part of the feature itself. Moreover, as the feature does not maintain a constant shape and orientation to the observer, the background subtraction will remove different portion of the feature at different parts along the loop, possibly interfering with the desired measurements. \p

Our results do not specifically rule out models that include spatially compact loops, but they do show that the more general veil-like case must be considered and explicitly ruled out before a given active region loop may be considered as a compact structure. Because veil-like structures are a more general, higher entropy class of shapes than isolated compact strands, and because current observations---including spectral filling-factor analysis---do not appear to rule out the veil case, it must be considered the most plausible scenario for any given observation, and interpretations that require compact form should be defended with compelling observational evidence.\p

Our results may bear importance to other areas of astrophysics and plasma physics. Our hypothesis that emitting plasma may be structured in veil-like manifolds \textit{and that it is difficult to tell if this is the case} may be relevant to more than just solar plasma structures, but to other magnetically confined hot plasmas as well.\p

\acknowledgements
This work was sponsored by NASA awards HGCR/NNX14AI14G and NNX16AG98G/S005. The work of JAK was supported by the NASA Heliophysics Guest Investigator program (NNX16AG98G) and the Internal Scientist Funding Model (competitive work package) program at GSFC.

\appendix
\section{Additional Examples of Wrinkle-Like Features}\label{sec_appendix}
In this appendix, we show several more examples of loops which appear to be projection artifacts akin to the feature studied in \secref{sec_single_loop}. This is not an extensive list, and it's purpose is not to make quantitative analysis of the fraction of loops which are projection artifacts. Such analysis needs to be a subject of a separate, large study. In this appendix, we instead demonstrate that the projections artifacts are ubiquitous enough that \textit{a} particular loop cannot \textit{a priori} be assumed to be a thin isolated structure in the volume. \p

The loops here are displayed in the manner similar to that from Figures~(\ref{loop_pos}) and~(\ref{loop_slices}): in each pair of figures, the first figure shows line-of-sight integrated image, an identified loop and slices across it, while the next figure shows structures in the volume which contribute to emission along the loop. Like in \figref{loop_slices}, we attempt to isolate subvolumes with bright features; unlike in \figref{loop_slices}, we usually plot LOS integral for more than one subvolume. The individual subvolumes are marked by colored bars above each slice image, and the corresponding colors on the LOS integrals mark the results for these subvolumes. While several subvolumes could contain a peak roughly corresponding to a given loop, note that in most cases, only one of these subvolumes has the peak at the right location \textit{on all slices}. \p

\begin{figure}[h]
  \begin{center} 
   \includegraphics[width=10cm]{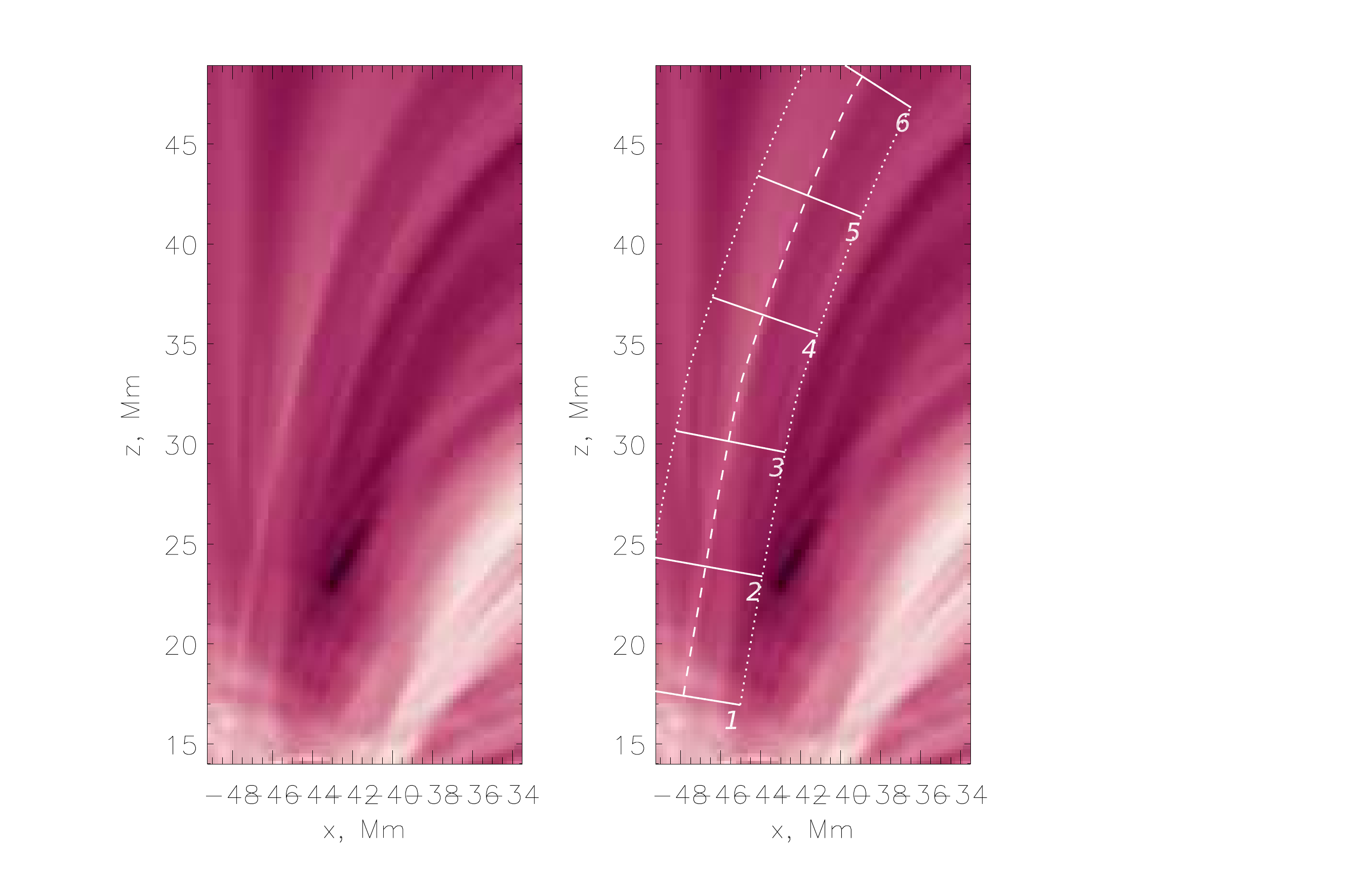}  
   \end{center} 
 \caption{Loop ``A01'' from the synthetic 211\AA~image, displayed in the same fashion as the one in \figref{loop_pos}. \textit{Left panel:} subregion in the synthetic 211\AA~image which shows a relatively clear and isolated loop. \textit{Right panel:} the traced loop, the slab around it and the slices across the loop which we further examine in the next figure.}
 \label{loop_a01_fov}
\end{figure}

\begin{figure}[h]
  \begin{center} 
   \includegraphics[width=27cm]{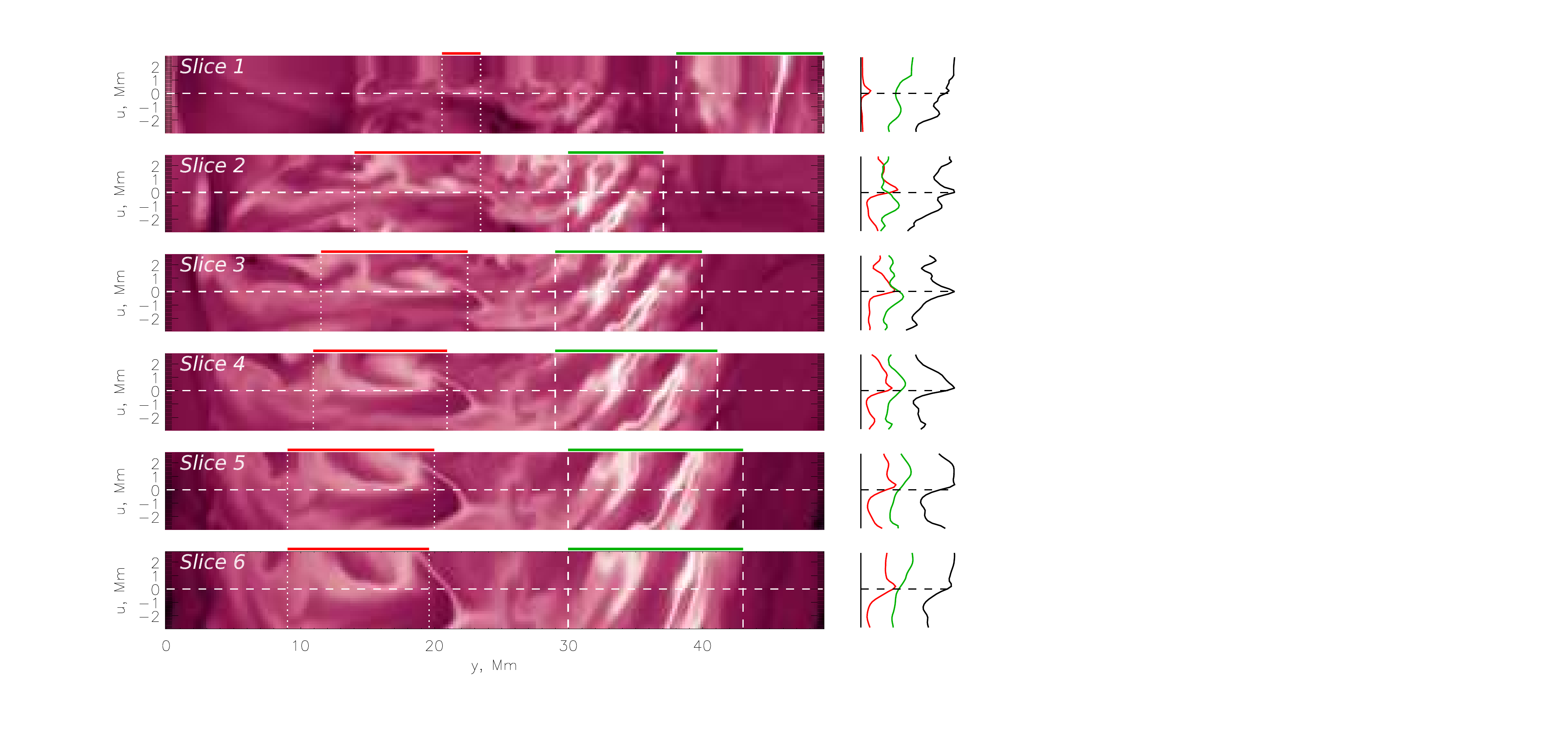}  
   \end{center} 
 \caption{Slices of volumetric emissivity across loop A01 from \figref{loop_a01_fov}, displayed in a similar fashion as \figref{loop_slices}. The left panels show slices in the volume which project into the lines 1-6 shown in \figref{loop_a01_fov}. Unlike in \figref{loop_slices}, which showed an integral over one subdomain, we here choose several subdomains with bright features which can project into the loop. The respective slices are indicated by colorbars above the subdomains in the left panels, and their LOS integrals are plotted in corresponding colors in the right panels. It appears that, while the ``green domain'' features dominate the emission background, the loop itself (the bump in the middle) is likely caused by the feature in the ``red'' domain. Indeed, the LOS integral over the ``red'' feature has a bump which is correlated with the loop location ($u=0$) \textit{in all slices}. Notice also that a skewed background is in fact part of the ``red'' feature itself (although the other parts of the volume further amplify this effect).} 
 \label{loop_a01_slices}
\end{figure}

\begin{figure}[h]
  \begin{center} 
   \includegraphics[width=10cm]{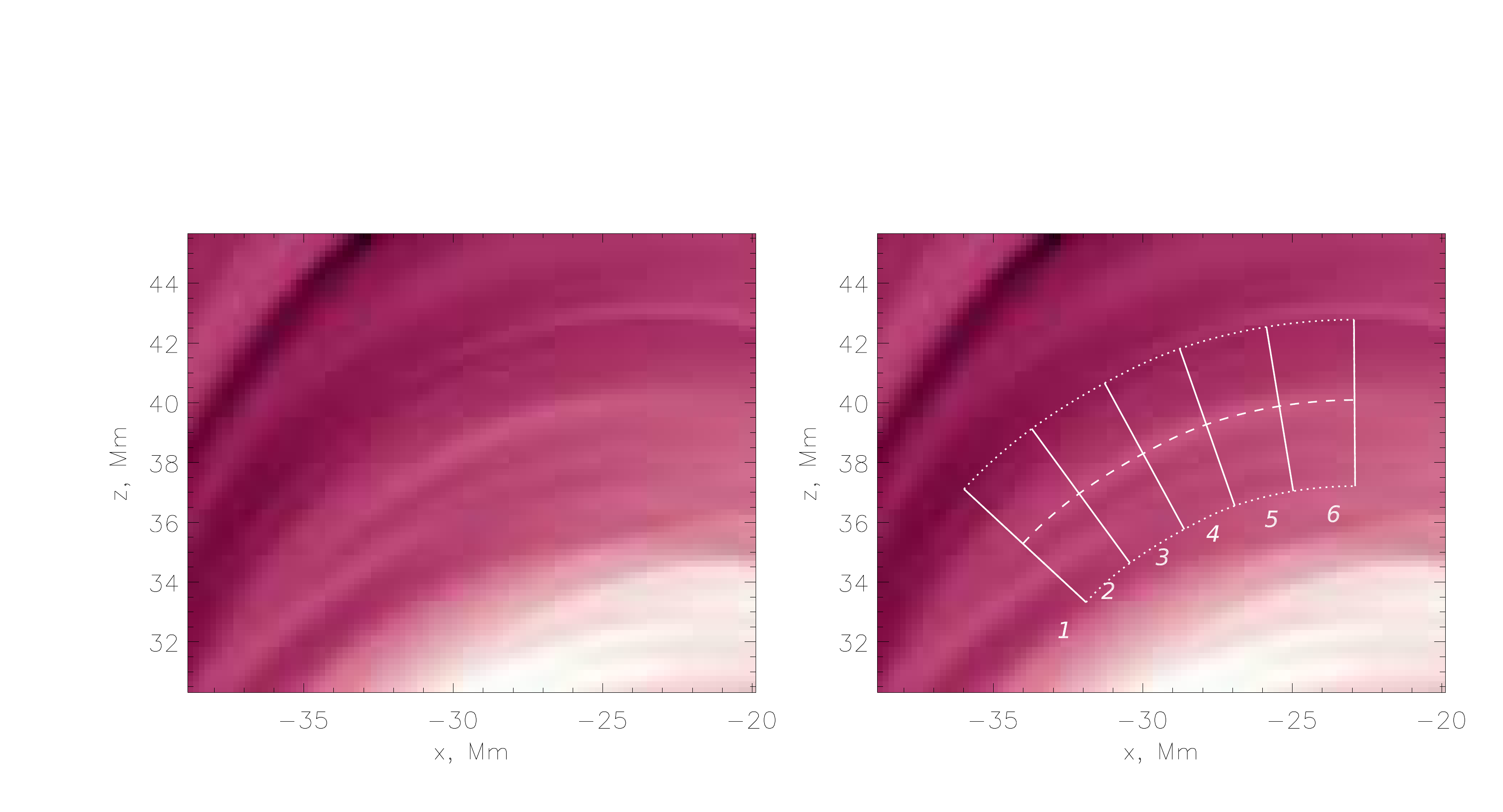}  
   \end{center} 
 \caption{Loop A02 from the synthetic 211\AA~image, displayed in the same fashion as the one in \figref{loop_a01_fov}.}
 \label{loop_a02_fov}
\end{figure}

\begin{figure}[h]
  \begin{center} 
   \includegraphics[width=27cm]{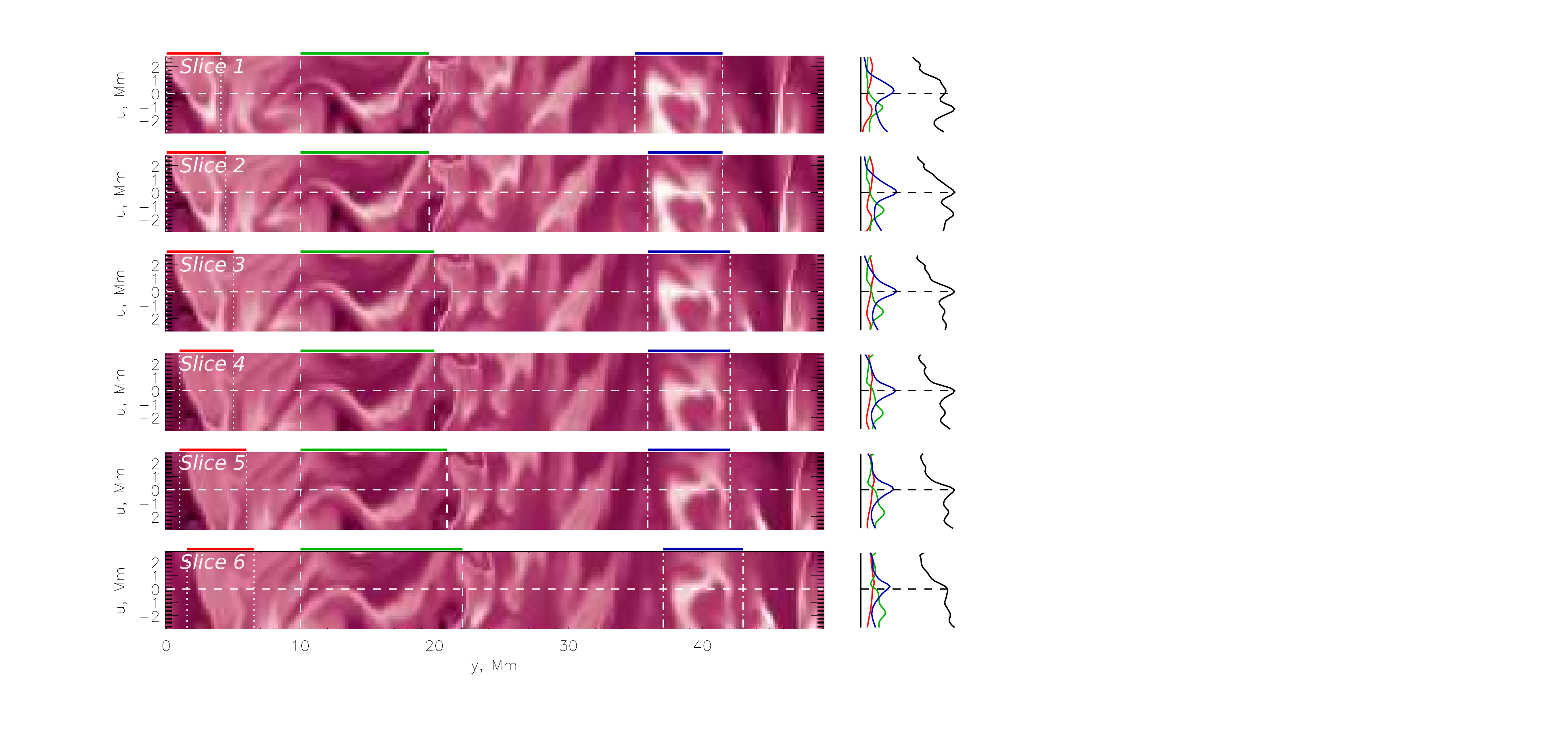}  
   \end{center} 
 \caption{Slices of volumetric emissivity across loop A02 from \figref{loop_a02_fov}, displayed in a similar fashion as \figref{loop_a01_slices}. The loop is likely produced by the feature in the ``blue'' domain.} 
 \label{loop_a02_slices}
\end{figure}

\begin{figure}[h]
  \begin{center} 
   \includegraphics[width=10cm]{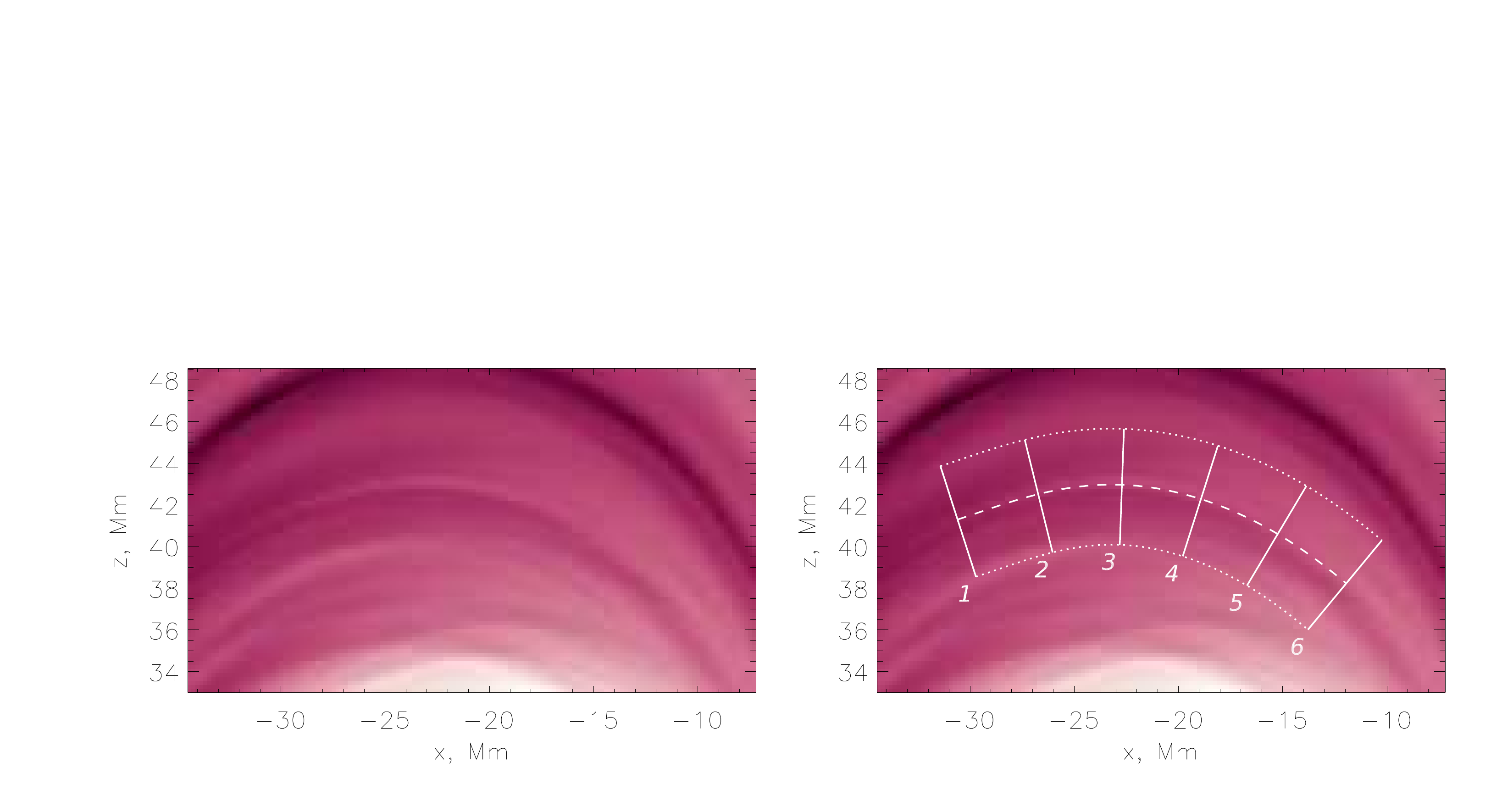}  
   \end{center} 
 \caption{Loop A03 from the synthetic 211\AA~image, displayed in the same fashion as the one in \figref{loop_a01_fov}.}
 \label{loop_a03_fov}
\end{figure}

\begin{figure}[h]
  \begin{center} 
   \includegraphics[width=27cm]{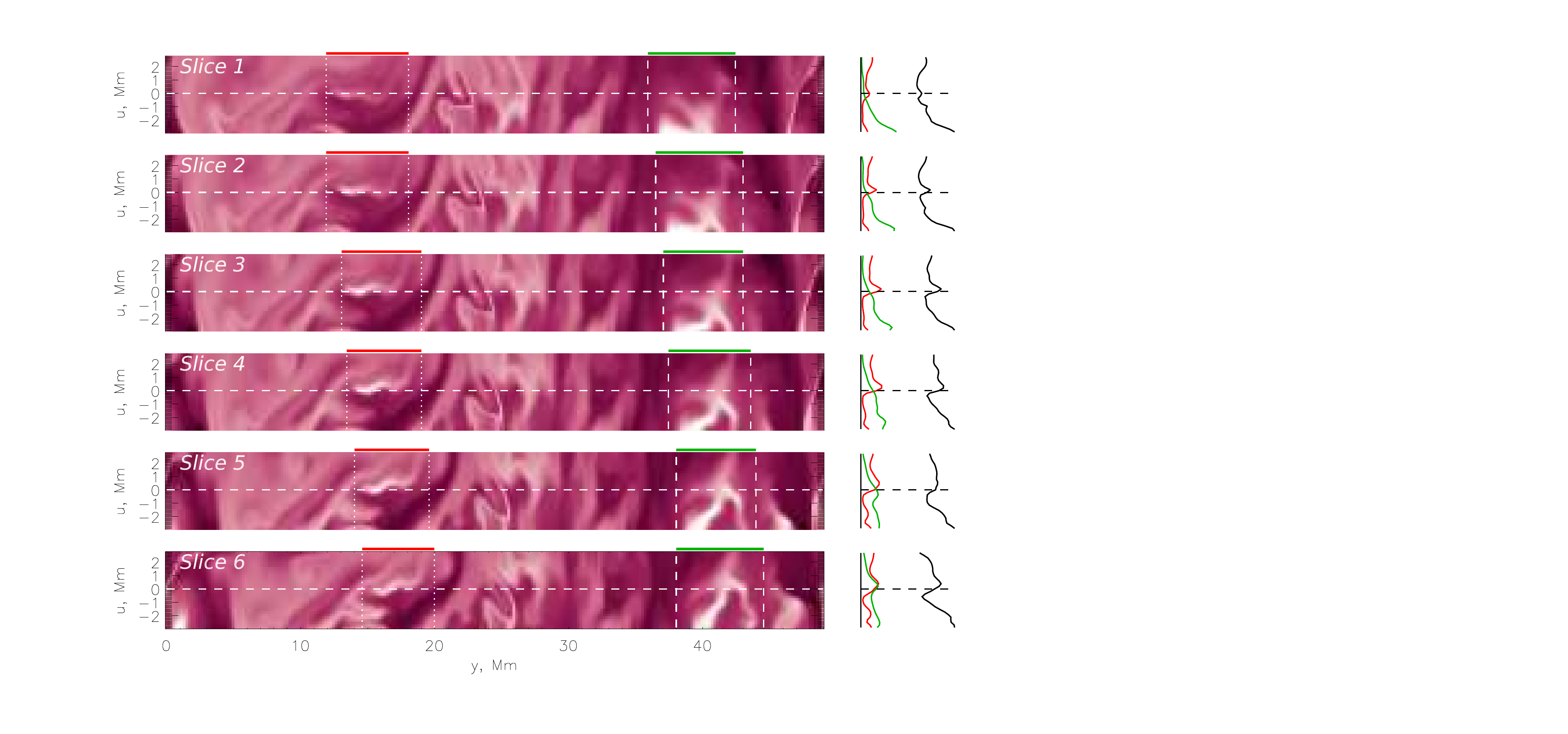}  
   \end{center} 
 \caption{Slices of volumetric emissivity across loop A03 from \figref{loop_a03_fov}, displayed in a similar fashion as \figref{loop_a01_slices}. The loop is likely produced by the feature in the ``red'' domain.} 
 \label{loop_a03_slices}
\end{figure}

\begin{figure}[h]
  \begin{center} 
   \includegraphics[width=10cm]{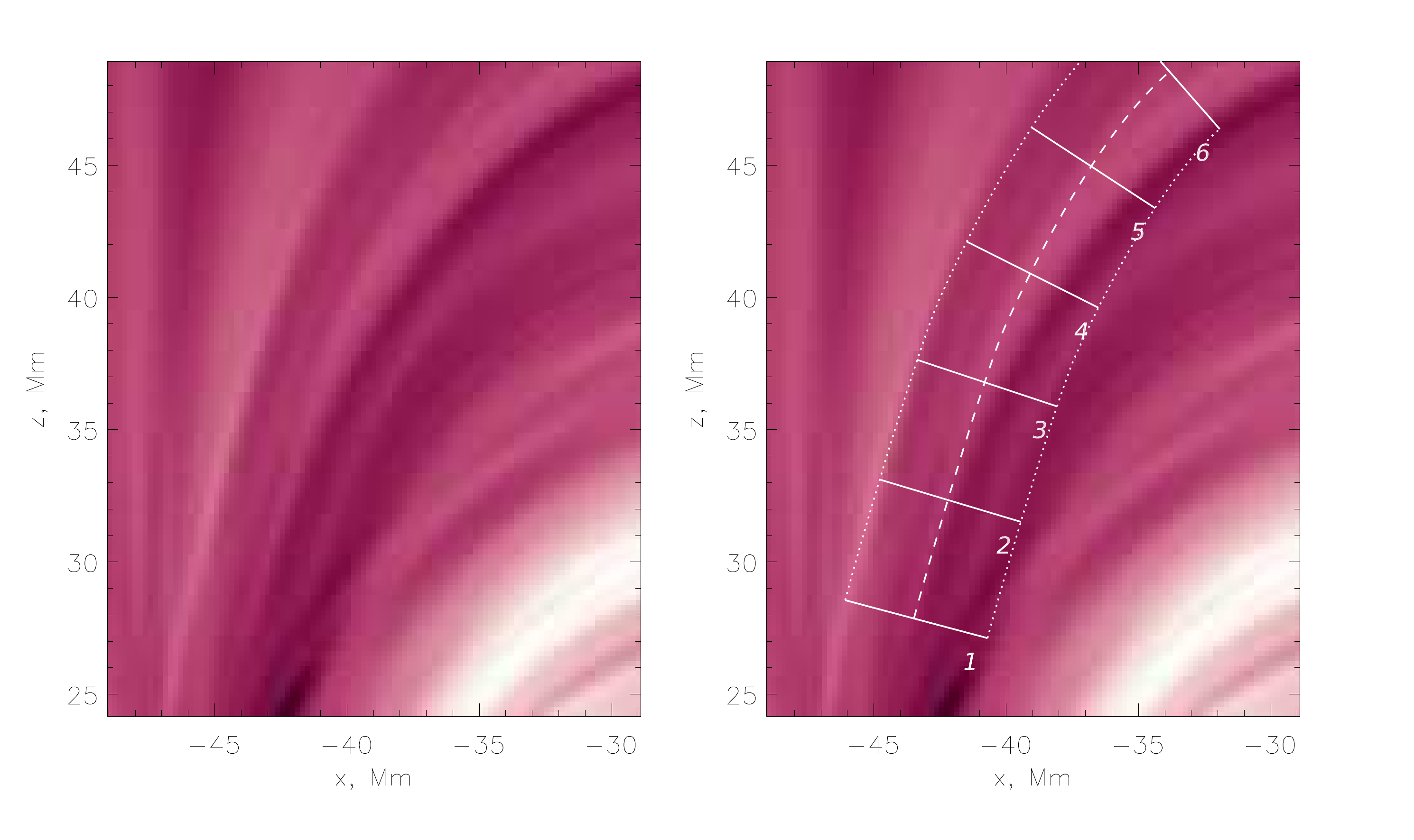}  
   \end{center} 
 \caption{Loop A04 from the synthetic 211\AA~image, displayed in the same fashion as the one in \figref{loop_a01_fov}.}
 \label{loop_a04_fov}
\end{figure}

\begin{figure}[h]
  \begin{center} 
   \includegraphics[width=27cm]{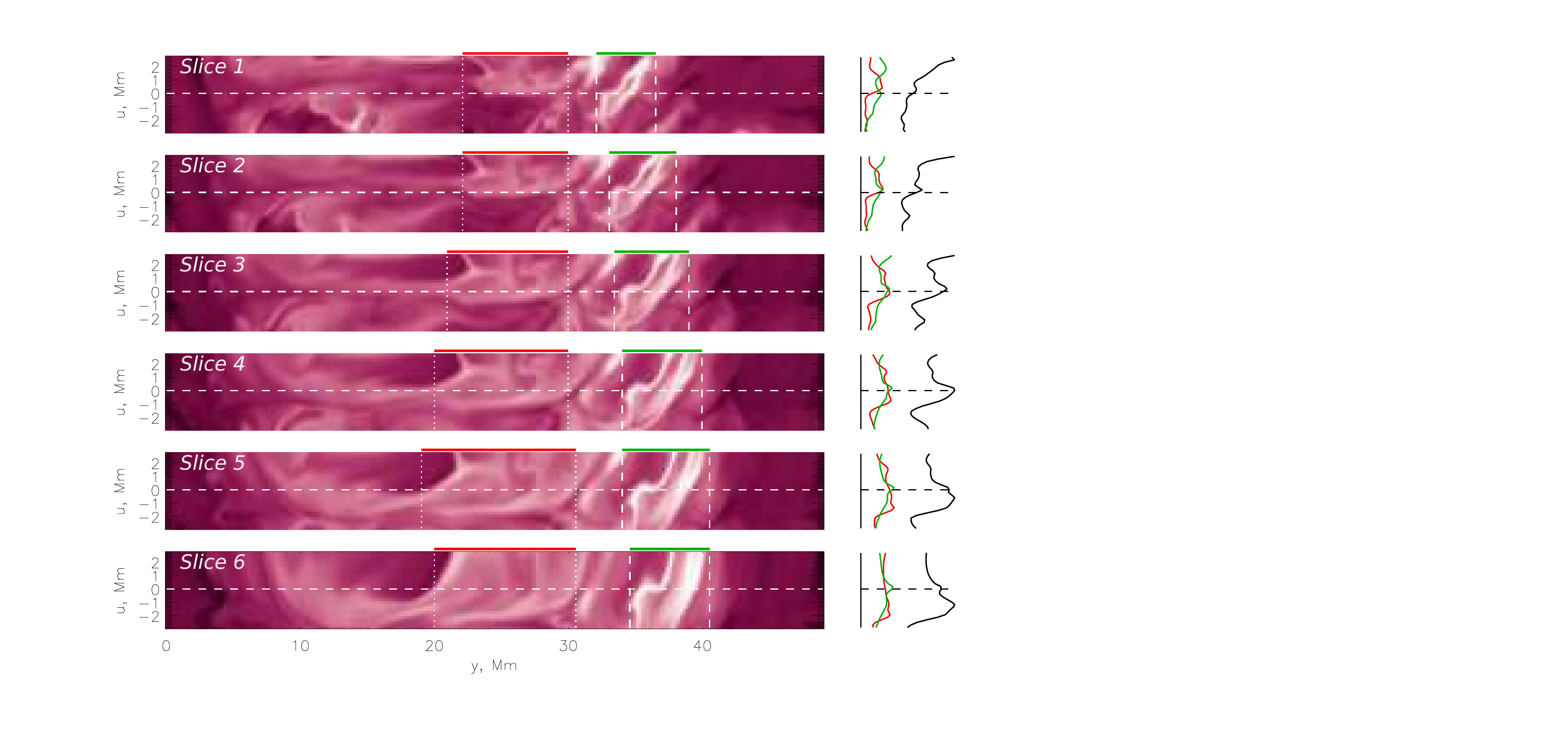}  
   \end{center} 
 \caption{Slices of volumetric emissivity across loop A04 from \figref{loop_a04_fov}, displayed in a similar fashion as \figref{loop_a01_slices}. The loop is likely produced by the feature in the ``green'' domain.} 
 \label{loop_a04_slices}
\end{figure}

\begin{figure}[h]
  \begin{center} 
   \includegraphics[width=10cm]{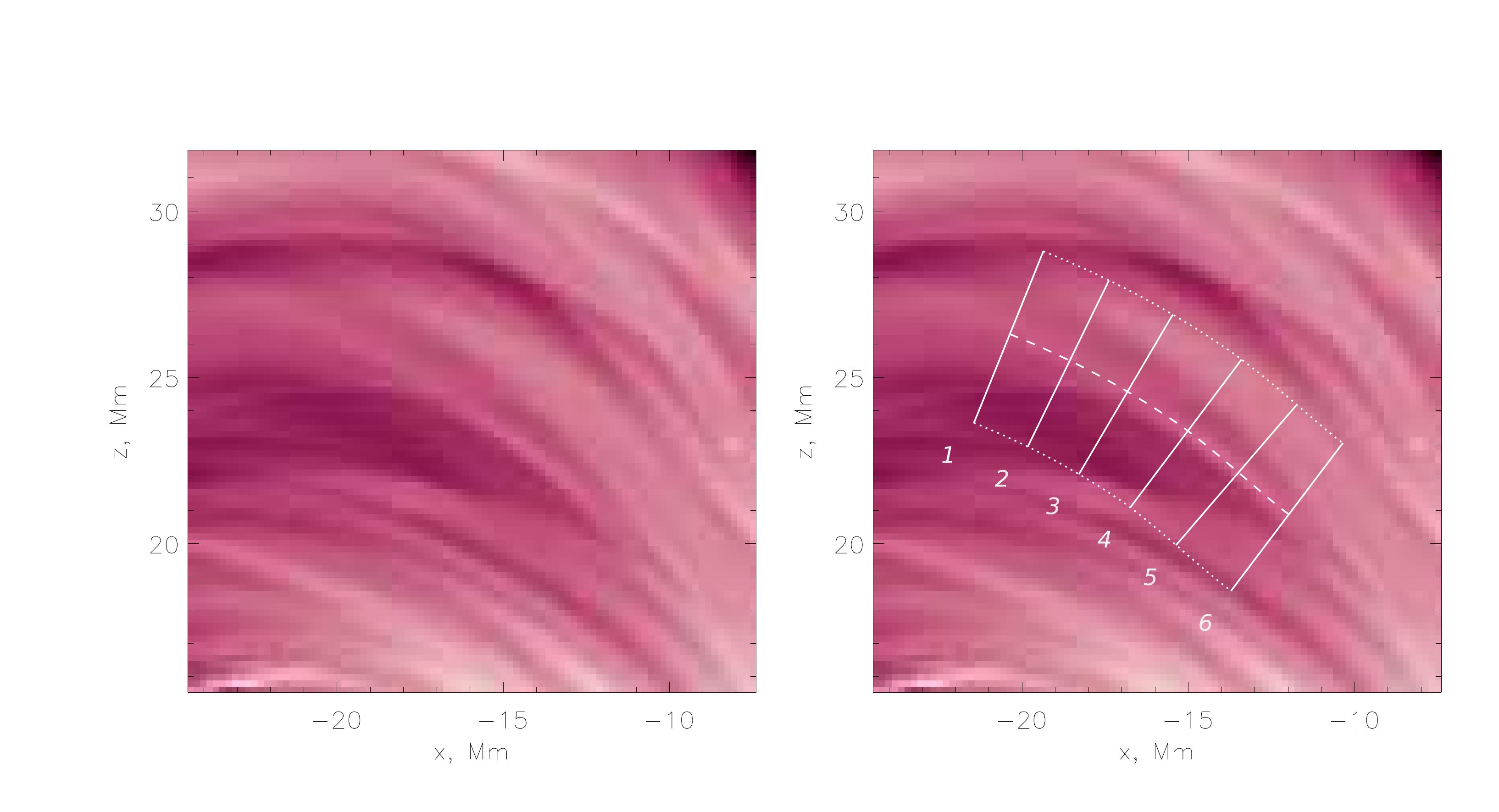}  
   \end{center} 
 \caption{Loop A05 from the synthetic 211\AA~image, displayed in the same fashion as the one in \figref{loop_a01_fov}.}
 \label{loop_a05_fov}
\end{figure}

\begin{figure}[h]
  \begin{center} 
   \includegraphics[width=27cm]{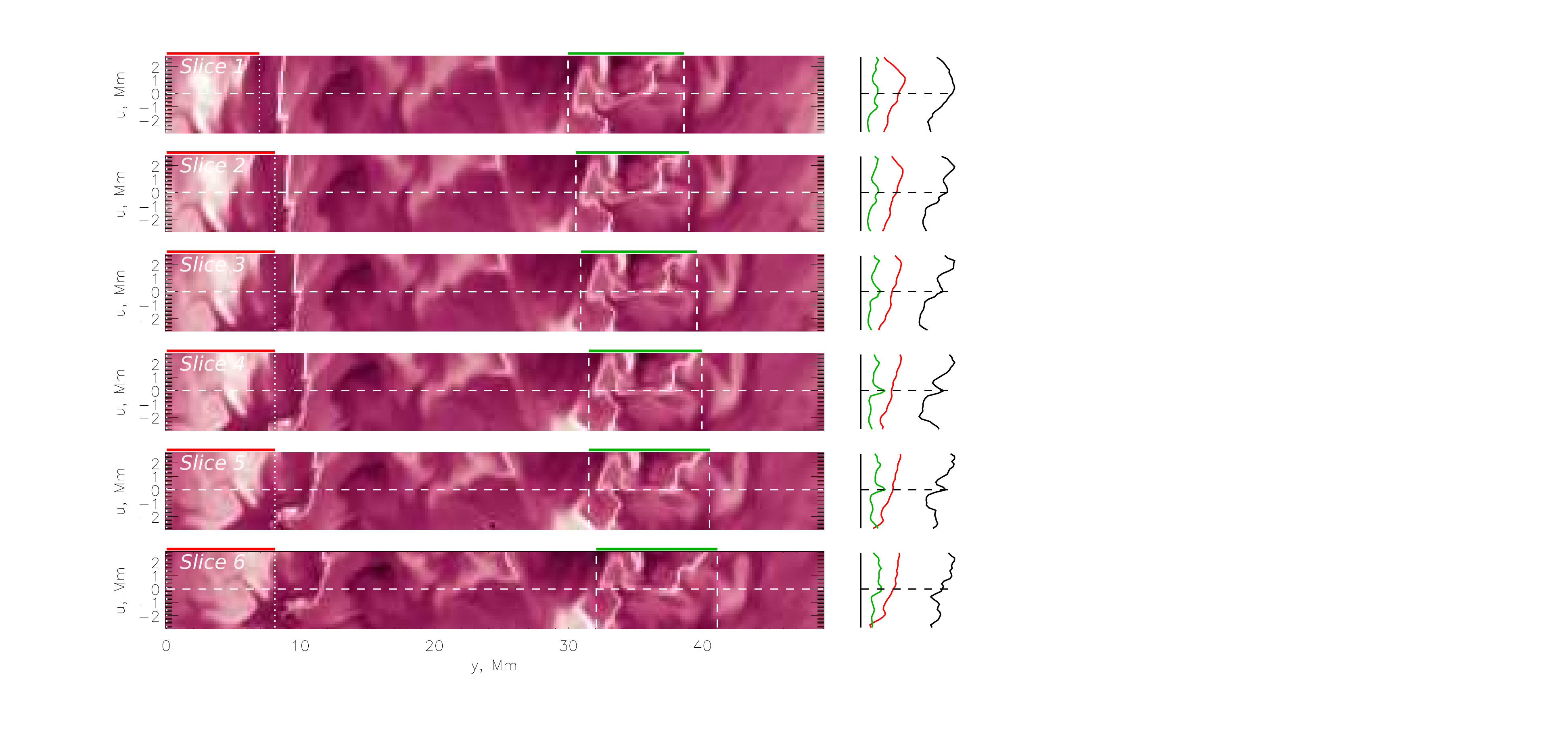}  
   \end{center} 
 \caption{Slices of volumetric emissivity across loop A05 from \figref{loop_a05_fov}, displayed in a similar fashion as \figref{loop_a01_slices}. The loop is likely produced by the feature in the ``green'' domain.} 
 \label{loop_a05_slices}
\end{figure}

\begin{figure}[h]
  \begin{center} 
   \includegraphics[width=10cm]{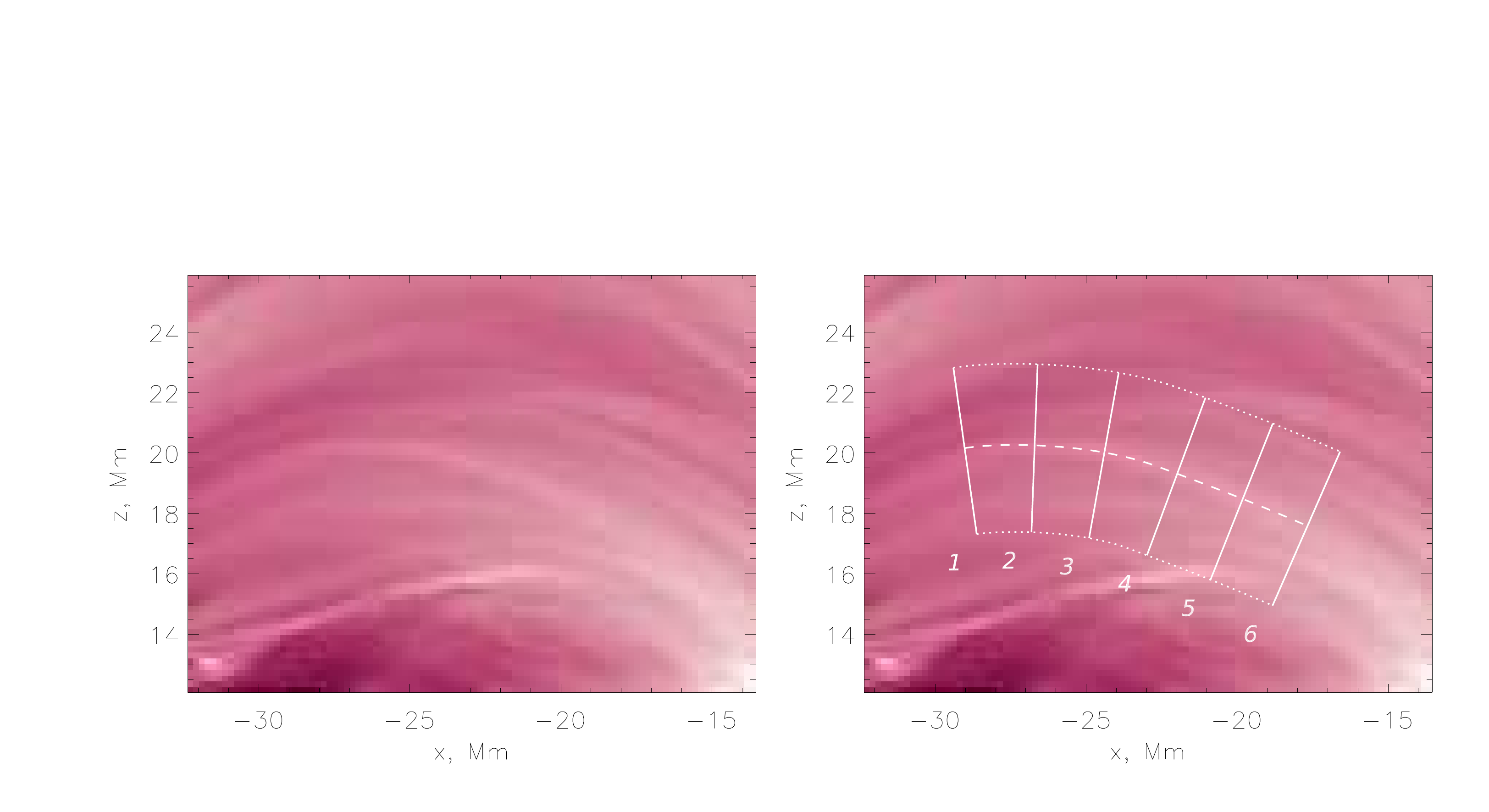}  
   \end{center} 
 \caption{Loop A06 from the synthetic 211\AA~image, displayed in the same fashion as the one in \figref{loop_a01_fov}.}
 \label{loop_a06_fov}
\end{figure}

\begin{figure}[h]
  \begin{center} 
   \includegraphics[width=27cm]{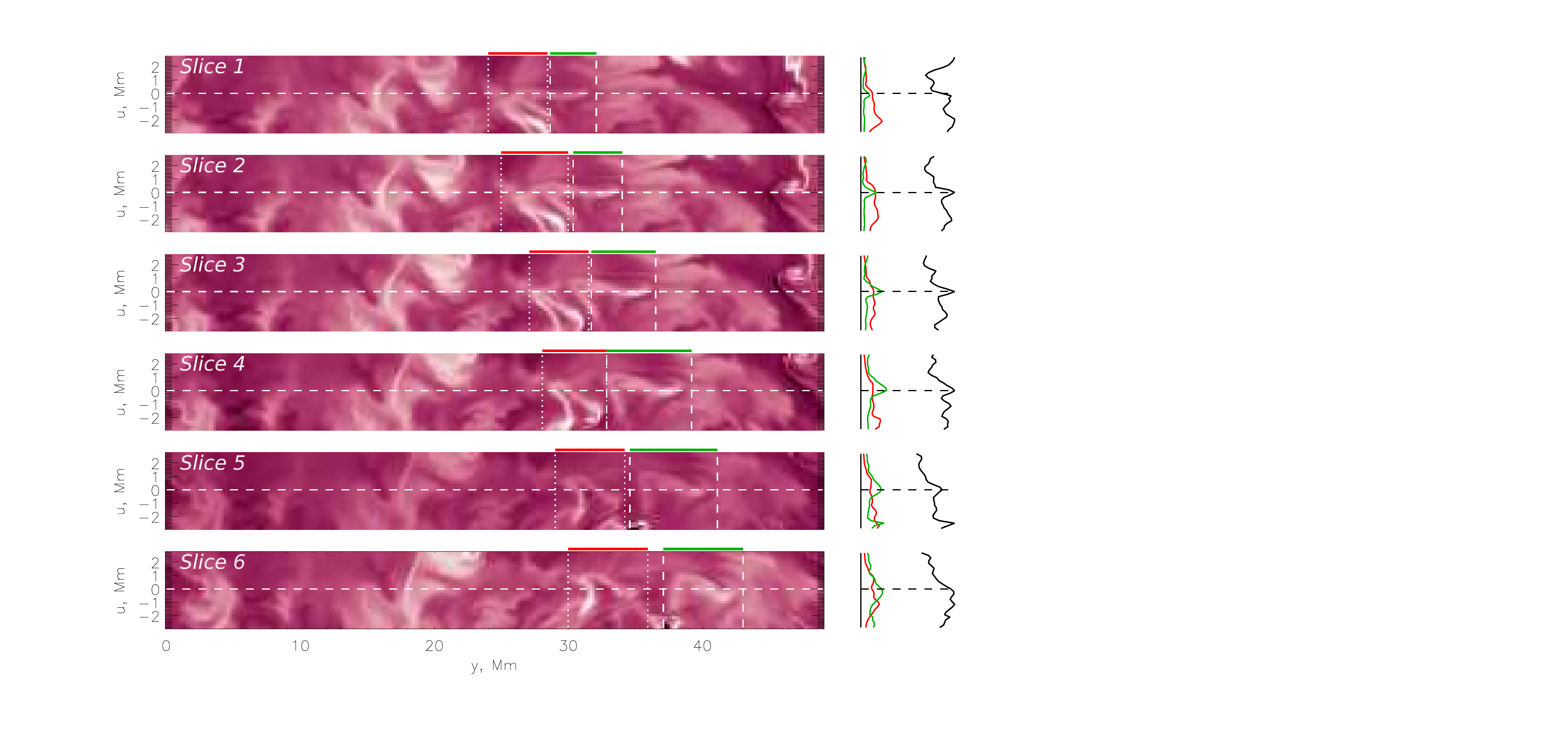}  
   \end{center} 
 \caption{Slices of volumetric emissivity across loop A06 from \figref{loop_a06_fov}, displayed in a similar fashion as \figref{loop_a01_slices}. The loop is likely produced by the feature in the ``green'' domain.} 
 \label{loop_a06_slices}
\end{figure}

\begin{figure}[h]
  \begin{center} 
   \includegraphics[width=10cm]{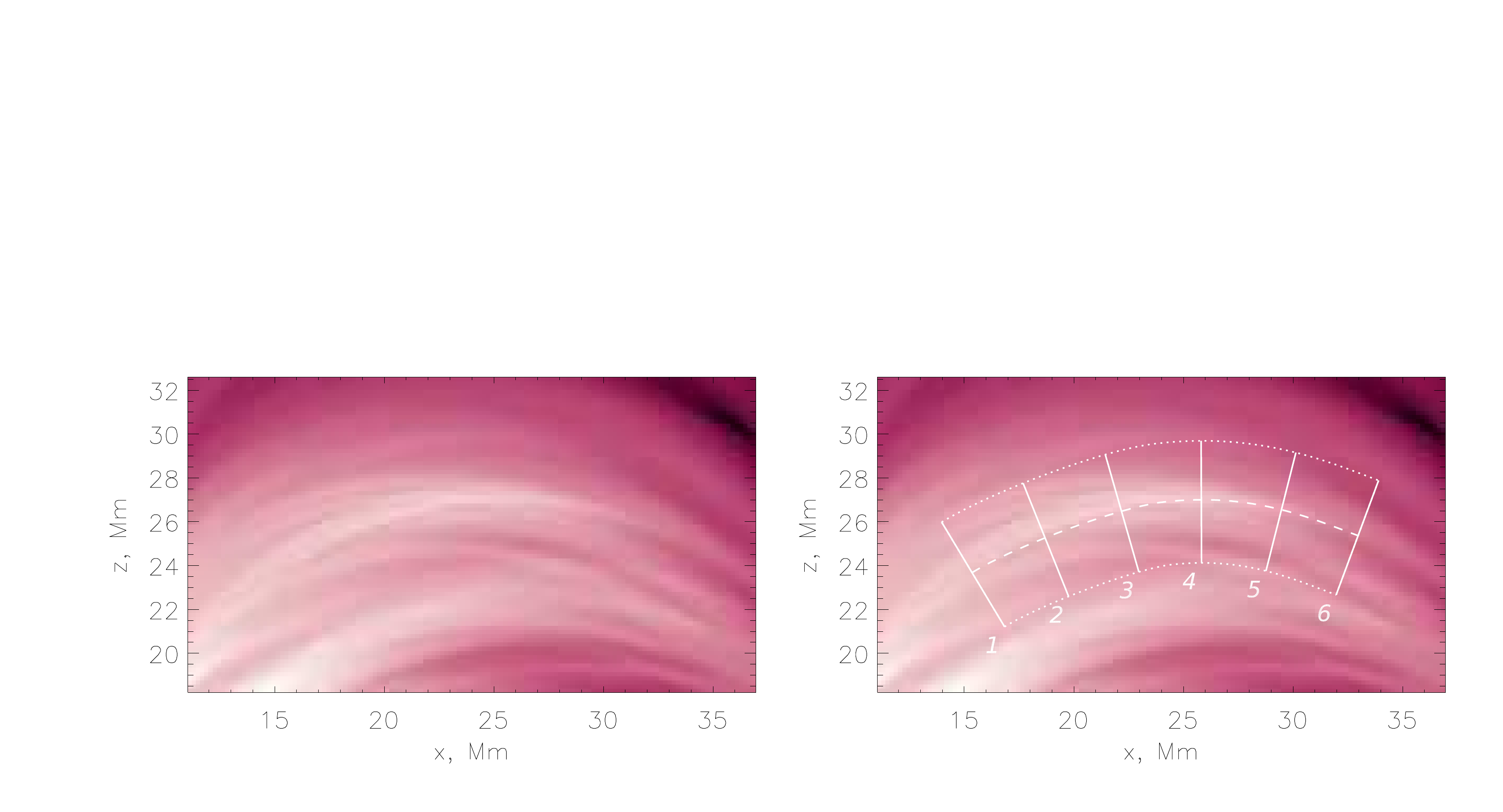}  
   \end{center} 
 \caption{Loop A07 from the synthetic 211\AA~image, displayed in the same fashion as the one in \figref{loop_a01_fov}.}
 \label{loop_a07_fov}
\end{figure}

\begin{figure}[h]
  \begin{center} 
   \includegraphics[width=27cm]{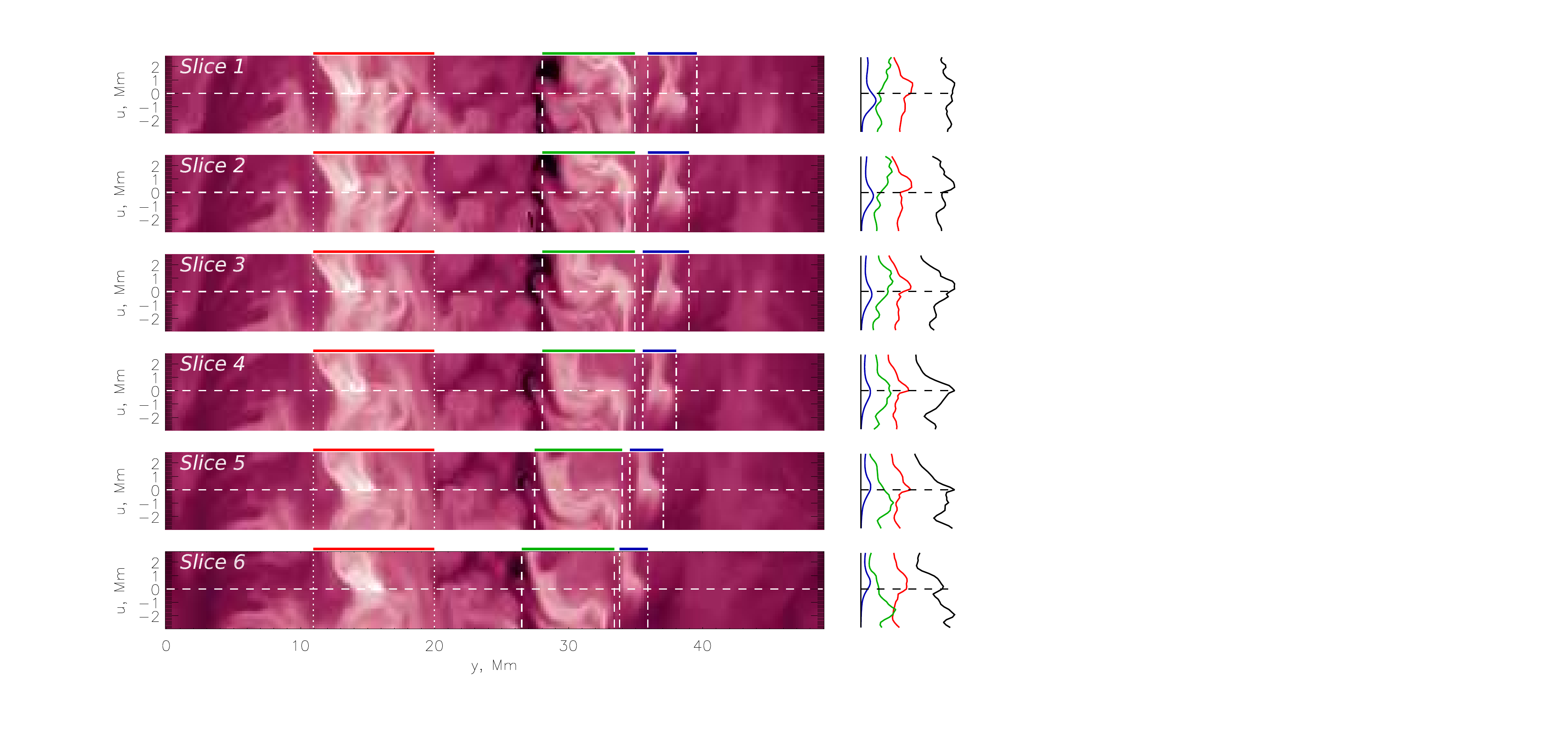}  
   \end{center} 
 \caption{Slices of volumetric emissivity across loop A07 from \figref{loop_a07_fov}, displayed in a similar fashion as \figref{loop_a01_slices}. The loop is likely produced by the feature in the ``red'' domain.} 
 \label{loop_a07_slices}
\end{figure}

\begin{figure}[h]
  \begin{center} 
   \includegraphics[width=10cm]{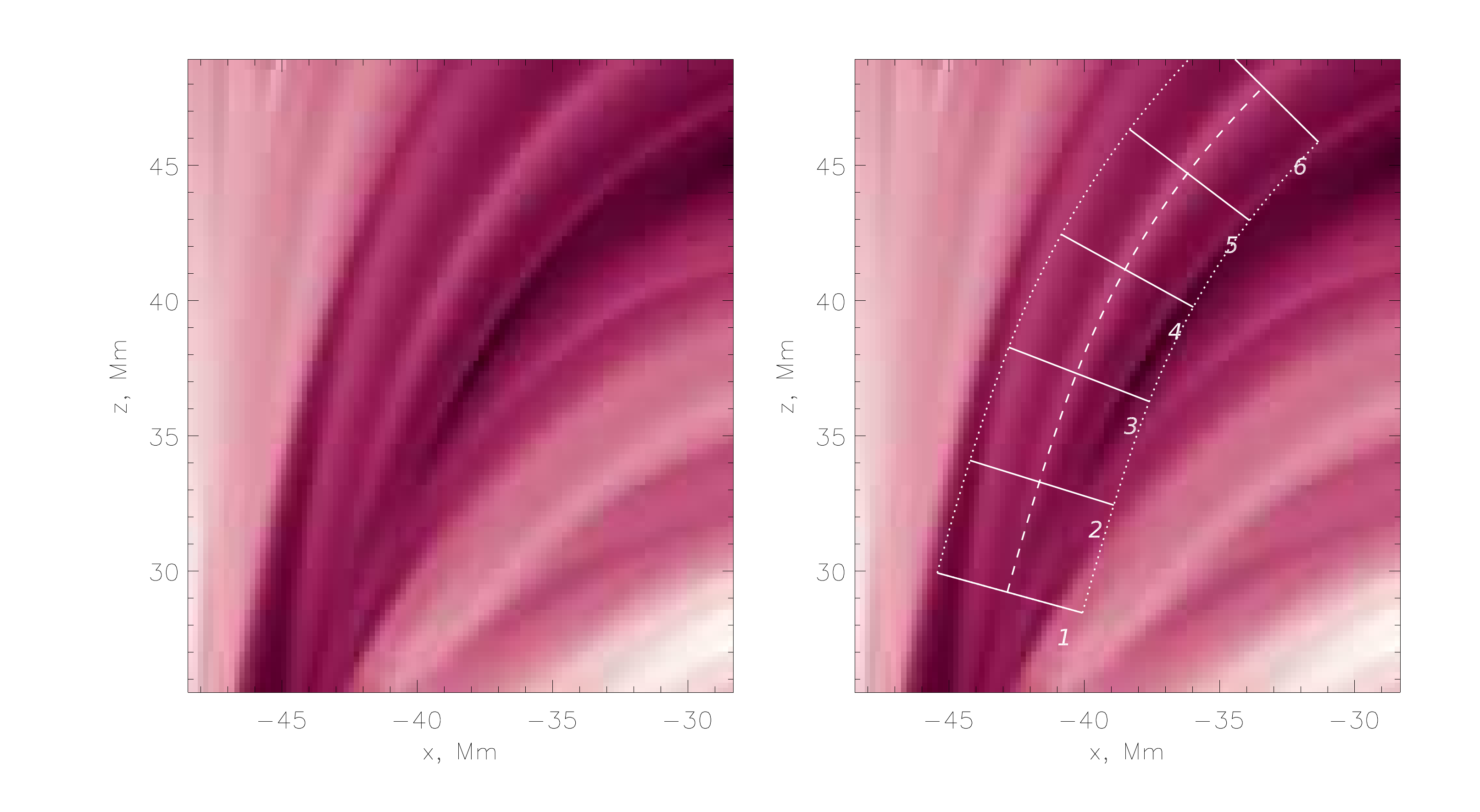}  
   \end{center} 
 \caption{Loop A8 from the synthetic 171\AA~image, displayed in the same fashion as the one in \figref{loop_a01_fov}.}
 \label{loop_a08_fov}
\end{figure}

\begin{figure}[h]
  \begin{center} 
   \includegraphics[width=27cm]{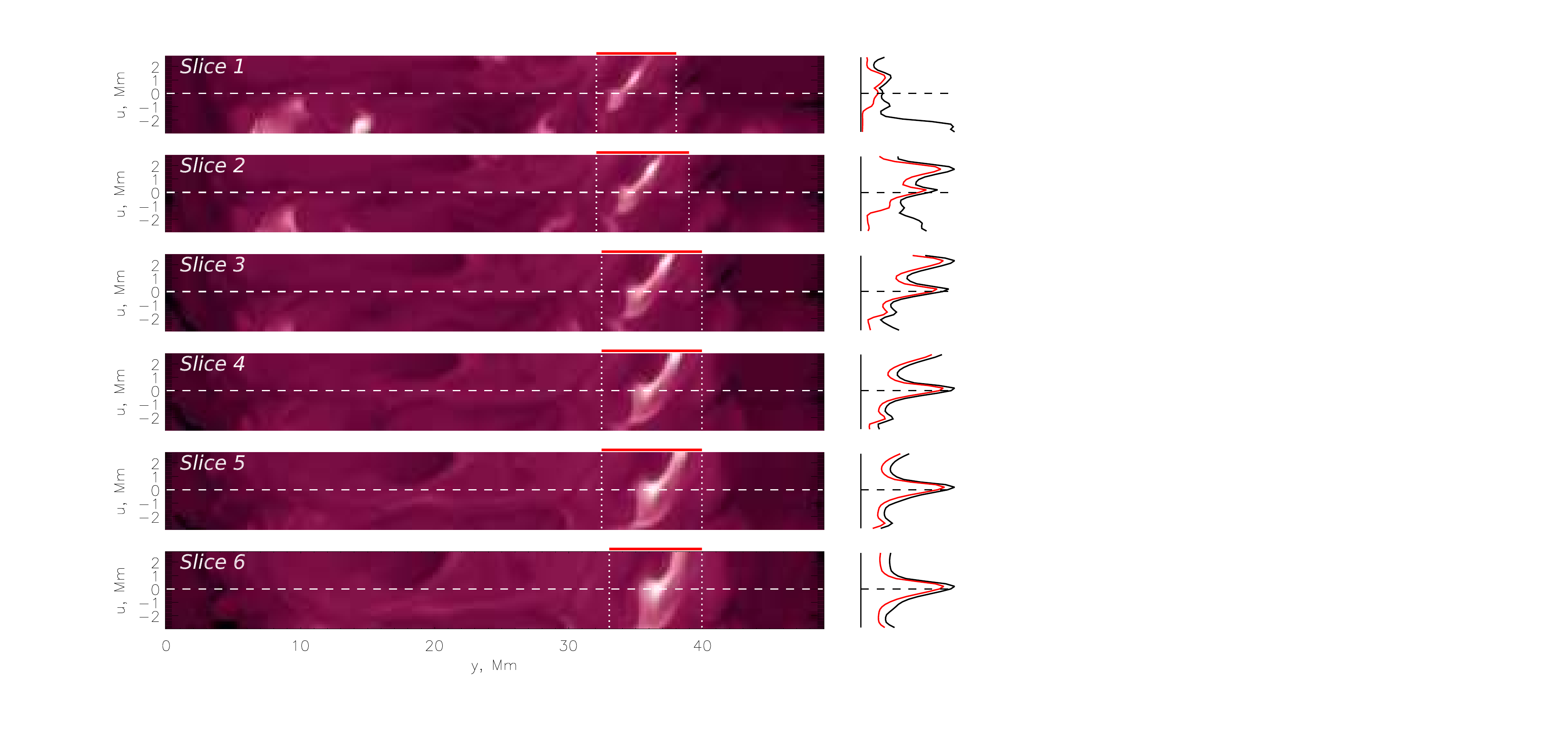}  
   \end{center} 
 \caption{Slices of volumetric emissivity across loop A8 from \figref{loop_a08_fov}, displayed in a similar fashion as \figref{loop_a01_slices}. The loop is likely produced by the feature in the ``red'' domain.} 
 \label{loop_a08_slices}
\end{figure}

\begin{figure}[h]
  \begin{center} 
   \includegraphics[width=10cm]{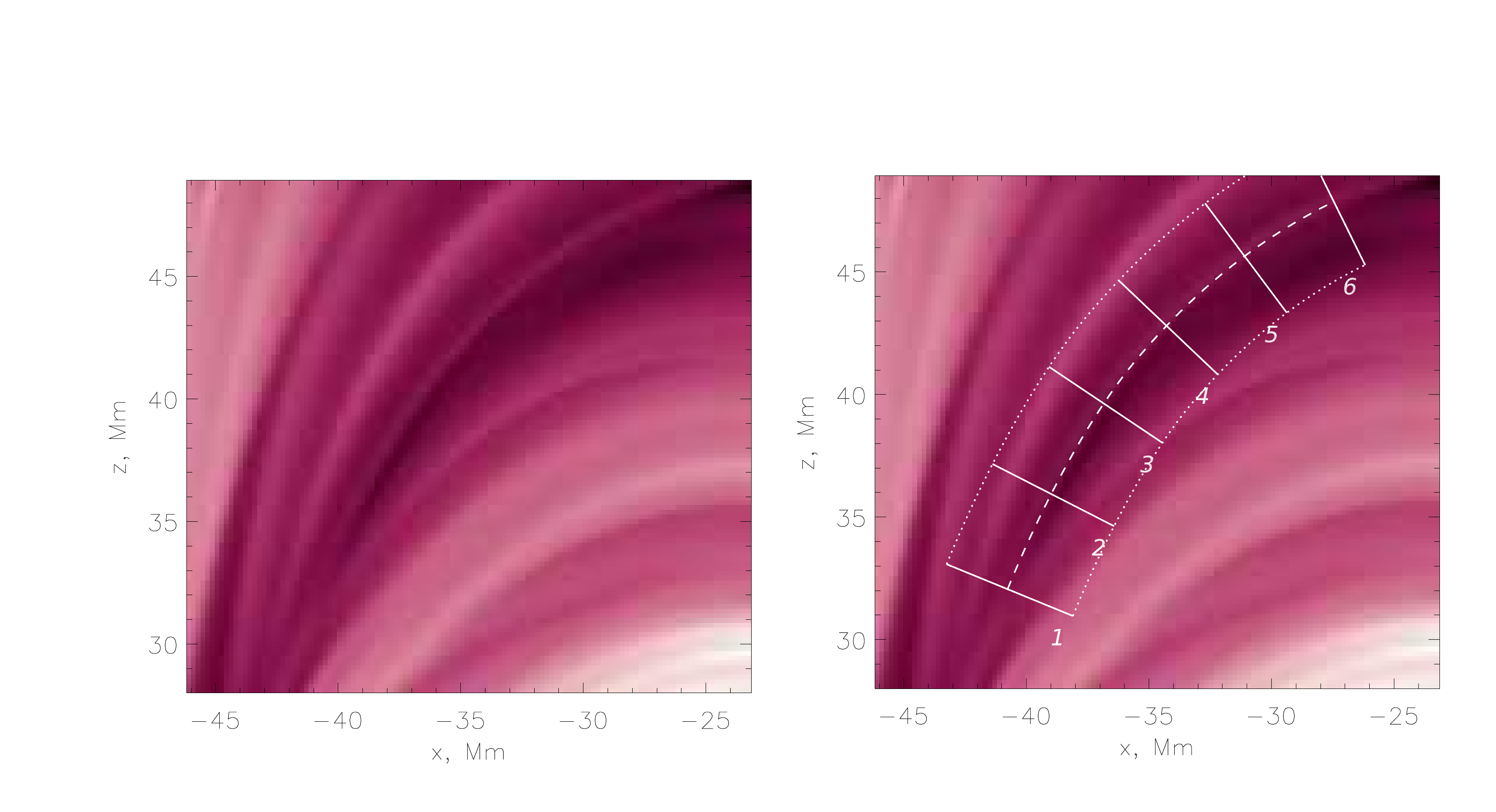}  
   \end{center} 
 \caption{Loop A09 from the synthetic 171\AA~image, displayed in the same fashion as the one in \figref{loop_a01_fov}.}
 \label{loop_a09_fov}
\end{figure}

\begin{figure}[h]
  \begin{center} 
   \includegraphics[width=27cm]{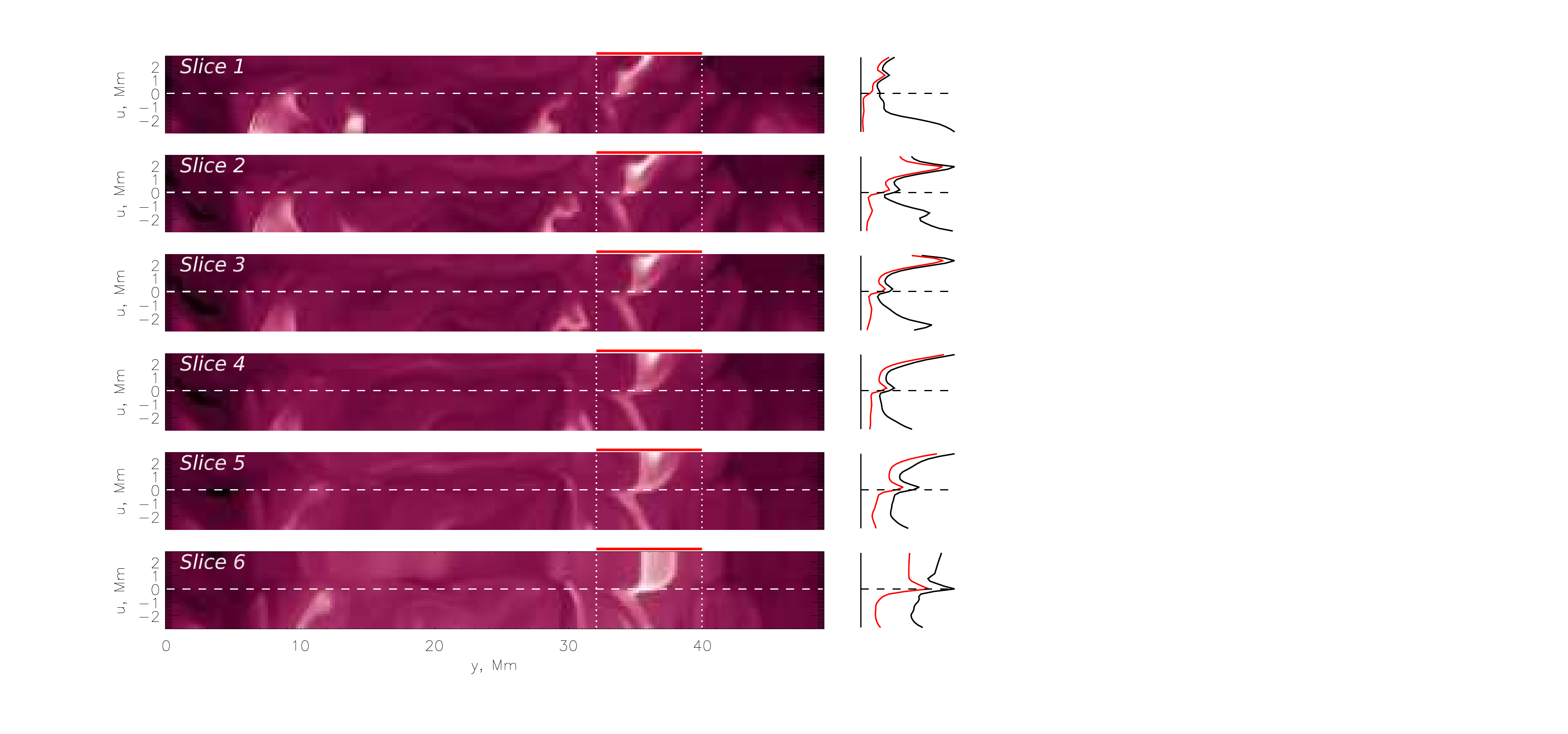}  
   \end{center} 
 \caption{Slices of volumetric emissivity across loop A9 from \figref{loop_a09_fov}, displayed in a similar fashion as \figref{loop_a01_slices}. The loop is likely produced by the feature in the ``red'' domain.} 
 \label{loop_a09_slices}
\end{figure}

\begin{figure}[h]
  \begin{center} 
   \includegraphics[width=10cm]{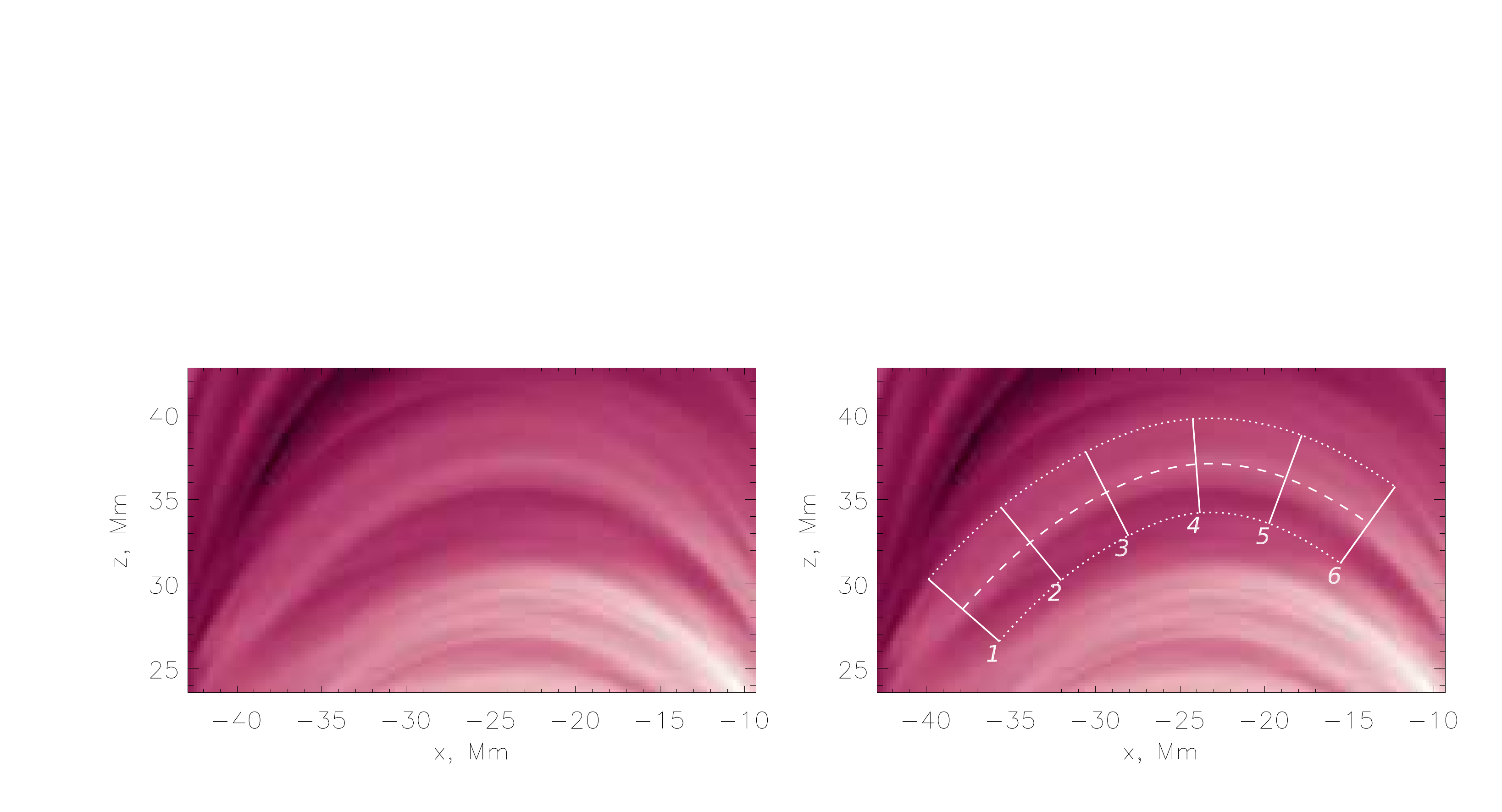}  
   \end{center} 
 \caption{Loop A10 from the synthetic 171\AA~image, displayed in the same fashion as the one in \figref{loop_a01_fov}.}
 \label{loop_a10_fov}
\end{figure}

\begin{figure}[h]
  \begin{center} 
   \includegraphics[width=27cm]{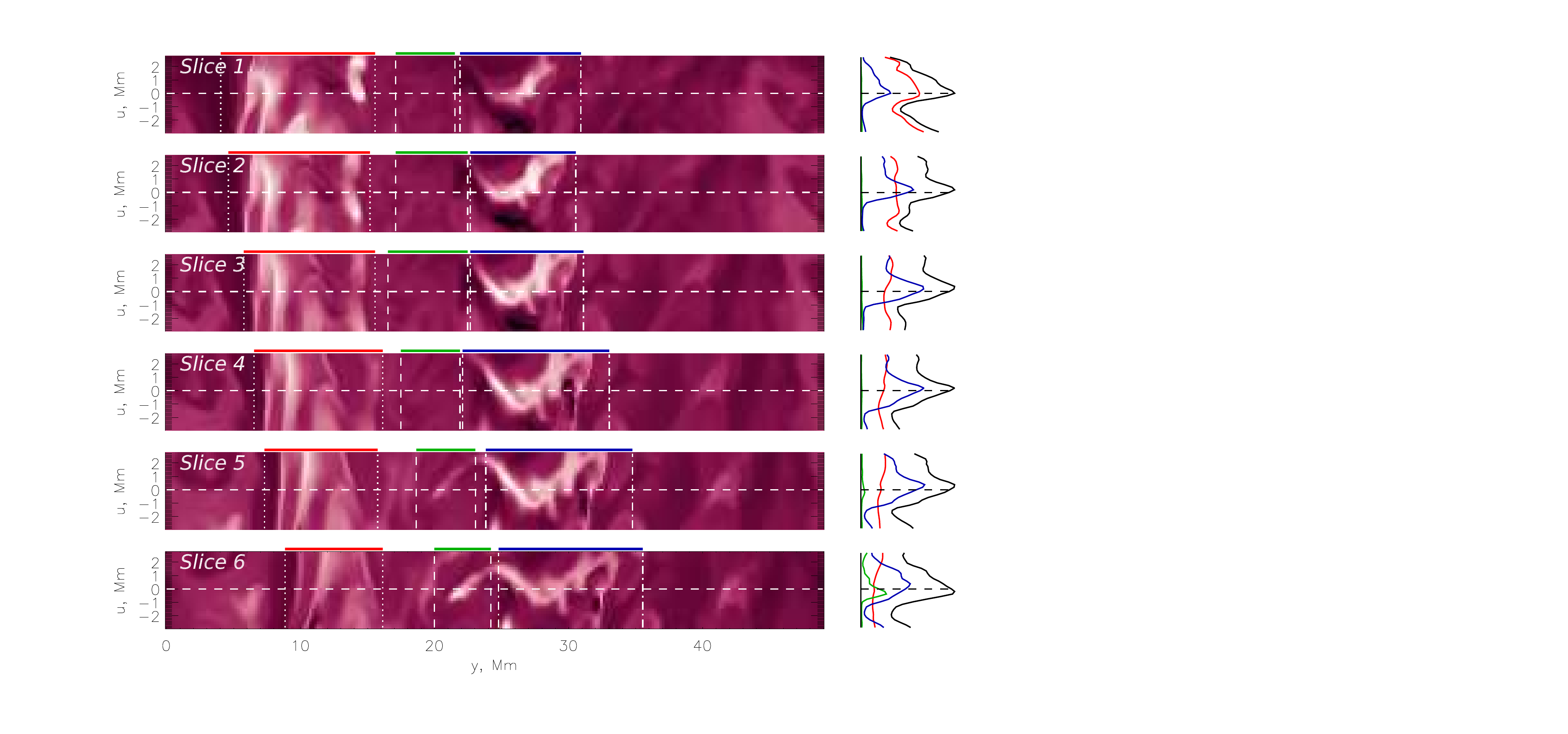}  
   \end{center} 
 \caption{Slices of volumetric emissivity across loop A10 from \figref{loop_a10_fov}, displayed in a similar fashion as \figref{loop_a01_slices}. The loop is likely produced by the feature in the ``blue'' domain.} 
 \label{loop_a10_slices}
\end{figure}

\begin{figure}[h]
  \begin{center} 
   \includegraphics[width=10cm]{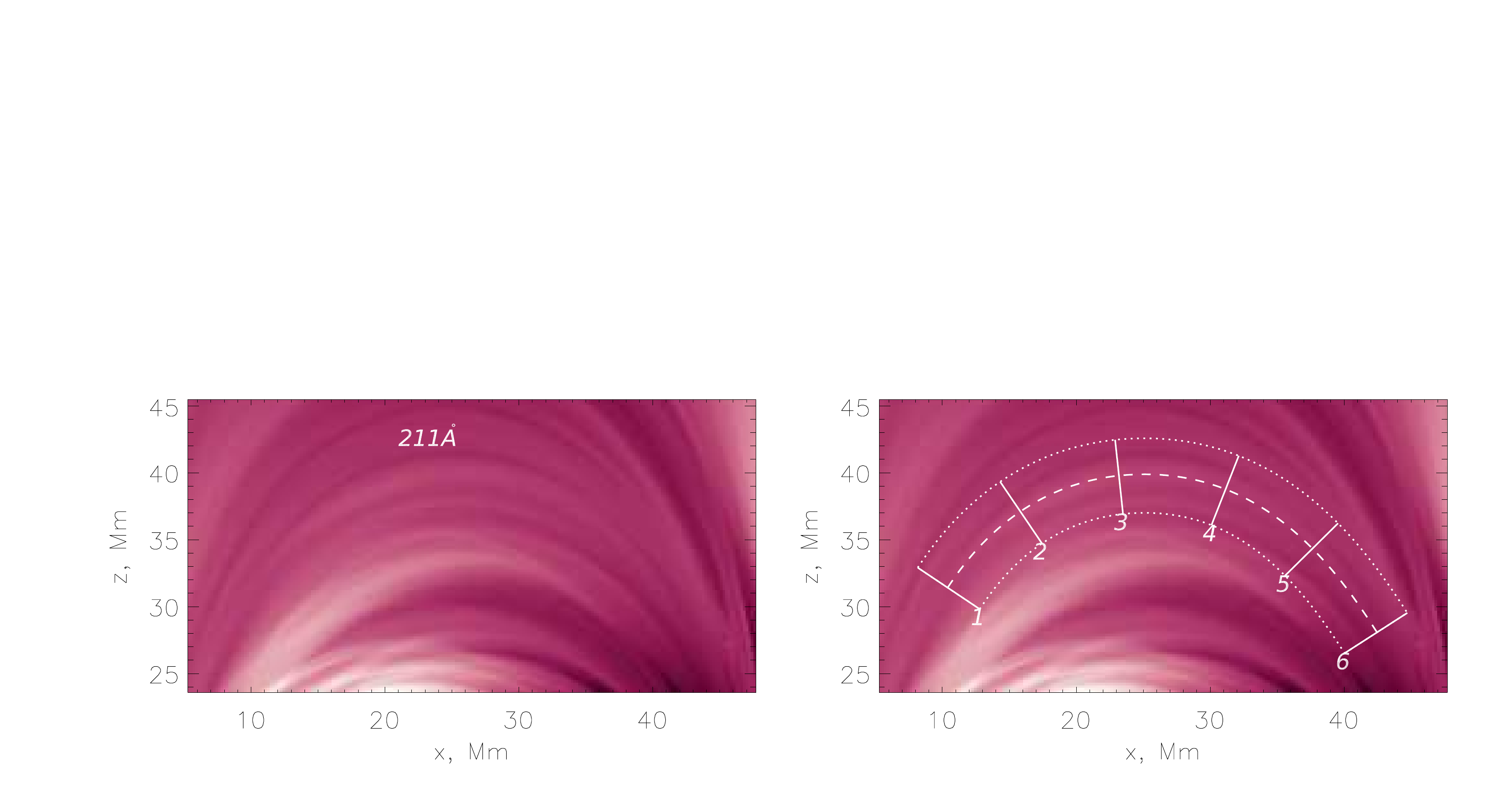}  
   \end{center} 
 \caption{Loop A11 from the synthetic 171\AA~image, displayed in the same fashion as the one in \figref{loop_a01_fov}.}
 \label{loop_a11_fov}
\end{figure}

\begin{figure}[h]
  \begin{center} 
   \includegraphics[width=27cm]{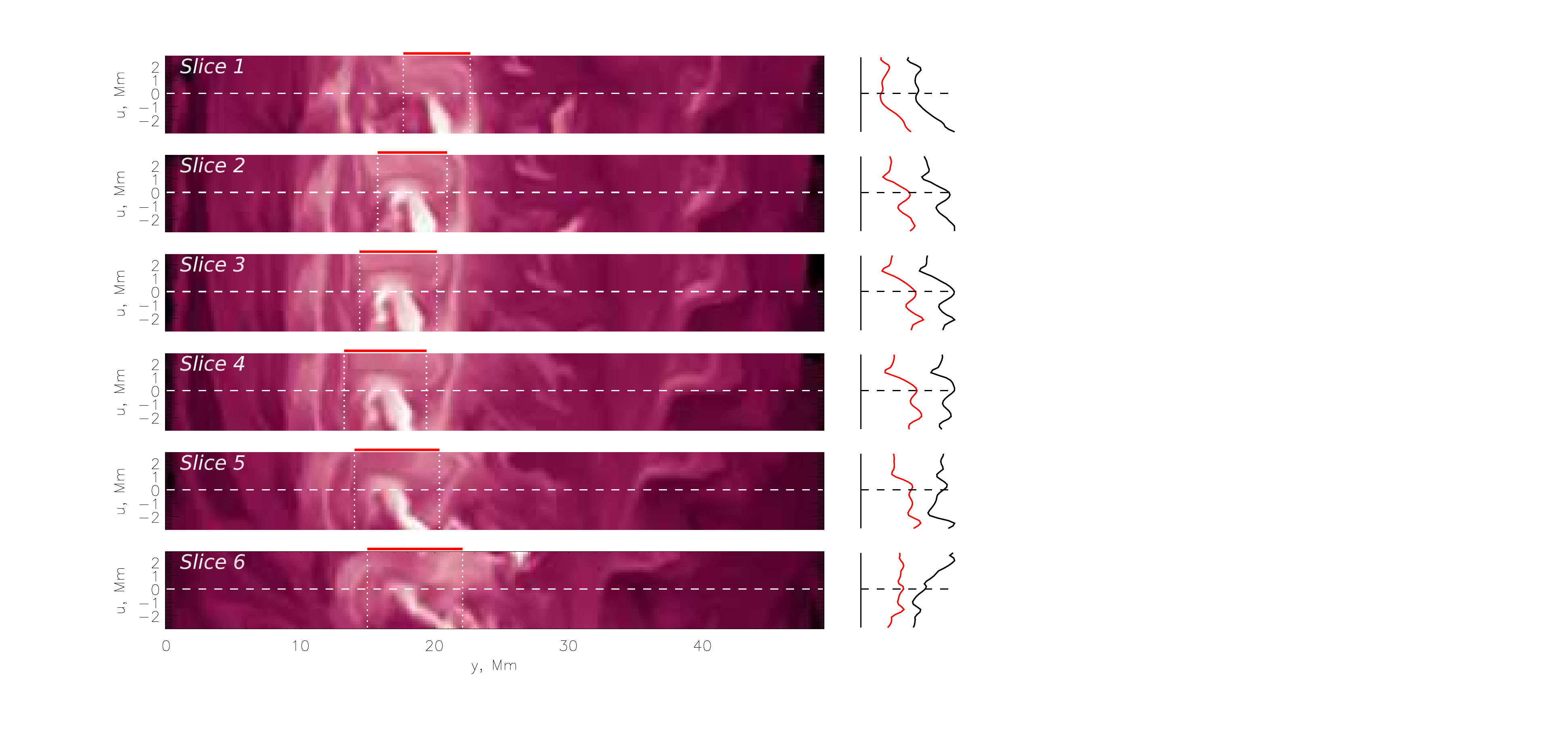}  
   \end{center} 
 \caption{Slices of volumetric emissivity across loop A11 from \figref{loop_a11_fov}, displayed in a similar fashion as \figref{loop_a01_slices}. The loop is likely produced by the feature in the ``red'' domain.} 
 \label{loop_a11_slices}
\end{figure}

\clearpage
\bibliography{apjmnemonic,short_abbrevs,total_lib,new_lib}

\end{document}